\documentclass[aps,prd,preprint,floats,epsf,superscriptaddress,nofootinbib,amsmath]{revtex4}
\usepackage[dvipdfmx]{graphicx}
\usepackage{amssymb,amsmath}
\usepackage{mathrsfs}
\usepackage{slashed}
\usepackage{indentfirst}
\usepackage{graphicx}
\usepackage{xcolor}
\usepackage{dsfont}
\usepackage{ulem}

\usepackage{cancel}

\usepackage[colorlinks,
            linkcolor=black,
            anchorcolor=black,
            citecolor=black
            ]{hyperref}

\begin{document}

\preprint{KUNS-2707}

\bigskip

\title{Model with a gauged lepton flavor SU(2) symmetry}

\author{Cheng-Wei Chiang}
\email[e-mail: ]{chengwei@phys.ntu.edu.tw}
\affiliation{Department of Physics, National Taiwan University, Taipei, Taiwan 10617, R.O.C.}
\affiliation{Institute of Physics, Academia Sinica, Taipei, Taiwan 11529, R.O.C.}
\affiliation{Kavli IPMU, University of Tokyo, Kashiwa, 277-8583, Japan}

\author{Koji Tsumura}
\email[e-mail: ]{ko2@gauge.scphys.kyoto-u.ac.jp}
\affiliation{Department of Physics, Kyoto University, Kyoto 606-8502, Japan}

\date{\today}

\begin{abstract}
We propose a model having a gauged $SU(2)$ symmetry associated with the second and third generations of leptons, dubbed $SU(2)_{\mu\tau}$, of which $U(1)_{\mathbb{L}_{\mu}-\mathbb{L}_{\tau}}$ is an Abelian subgroup.  In addition to the Standard Model fields, we introduce two types of scalar fields.  One exotic scalar field is an $SU(2)_{\mu\tau}$ doublet and SM singlet that develops a nonzero vacuum expectation value at presumably multi-TeV scale to completely break the $SU(2)_{\mu\tau}$ symmetry, rendering three massive gauge bosons.  At the same time, the other exotic scalar field, carrying electroweak as well as $SU(2)_{\mu\tau}$ charges, is induced to have a nonzero vacuum expectation value as well and breaks mass degeneracy between the muon and tau.  We examine how the new particles in the model contribute to the muon anomalous magnetic moment in the parameter space compliant with the Michel decays of tau.
\end{abstract}


\maketitle

\section{Introduction \label{sec:intro}}

A long-standing anomaly in flavor physics that has inspired pursues of physics beyond the Standard Model (SM) is the muon anomalous magnetic moment~\cite{Czarnecki:2001pv}.  A $3.6\sigma$ discrepancy between theory calculations within the SM and experimental data~\cite{Patrignani:2016xqp}:
\begin{align}
\Delta a_\mu \equiv a_\mu^{\rm exp} - a_\mu^{\rm SM}
= 288(63)(49)\times 10^{-11}
\label{eq:muondela}
\end{align}
has been reported\
\footnote{Other analyses give somewhat different estimates for the discrepancy, such as 
$\Delta a_\mu= (261 \pm 80) \times 10^{-11}$ in Ref.~\cite{Hagiwara:2011af}.  Ref.~\cite{Benayoun:2011mm} has some in-depth discussions about $\Delta a_\mu$ in various scenarios that mainly differ in the treatment of leading-order hadronic uncertainties and whether $\tau$ data are included.  In our numerical analysis, we use the result quoted in Eq.~\eqref{eq:muondela}.}.  One solution to this problem is to invoke a gauge boson with a sufficiently muon-philic coupling.  
Such a scenario can be realized in a gauged, anomaly-free $U(1)_{\mathbb{L}_{\mu}-\mathbb{L}_{\tau}}$ model, as proposed in Refs.~\cite{He:1990pn,Foot:1990mn,He:1991qd} and further examined in Refs.~\cite{Baek:2001kca,Ma:2001md}.  
It also features in the ability to explain the neutrino mixing data~\cite{Binetruy:1996cs,Bell:2000vh,Choubey:2004hn}.
The phenomenology of such a model at the LHC has been studied in Ref.~\cite{Harigaya:2013twa}.
It is recently analyzed that the $Z'$ boson with a mass in the range of ${\cal O}(1-100)$~MeV can explain $\Delta a_\mu$ with a fairly small coupling $\sim 2 \times (10^{-4} - 10^{-3})$, while being consistent with constraints such as the big bang nucleosynthesis bound on the effective number of extra neutrinos $\Delta N_{\rm eff} \le 1$~\cite{Ahlgren:2013wba} and the neutrino trident production~\cite{Altmannshofer:2014pba}.  The remaining parameter space can be soon probed or ruled out by future searches~\cite{Gninenko:2014pea}.

In this work, we propose a non-Abelian gauged flavor symmetry, $SU(2)_{\mu\tau}$, between the second and third generations of the SM leptons\
\footnote{This kind of extensions was briefly mentioned in some earlier works~\cite{He:1991qd,Heeck:2011wj}}.  After symmetry breaking induced by the vacuum expectation value (VEV) of a scalar $SU(2)_{\mu\tau}$ doublet, the gauged flavor group has three massive gauge bosons, denoted by $X_{3,\pm}$, with $X_3$ corresponding to the $U(1)_{\mathbb{L}_{\mu}-\mathbb{L}_{\tau}}$ subgroup and $X_{\pm}$ coupled with currents that exchange the muon and tau numbers by one unit.  To break the mass degeneracy between $\mu$ and $\tau$ as otherwise required by the new flavor symmetry, we further introduce a set of scalar fields that transform as a triplet under $SU(2)_{\mu\tau}$ and a doublet under the SM $SU(2)_L$.

An intriguing property of the model is the muon number conservation, respected by the new gauge interactions and Yukawa couplings of the muon and tau with the new scalar bosons.  As a consequence, no lepton flavor-violating processes are allowed in the model.  Another feature of the model is that no $Z$-$X_3$ mixing is allowed at tree level according to the proposed symmetry breaking pattern.

To comply with the current 125-GeV Higgs data, we will consider a nearly alignment limit of this model where the Higgs couplings to the SM particles are SM-like.  We compute the contributions of the new gauge bosons and scalar bosons to $\Delta a_\mu$ and the Michel decays of $\tau \to \ell \nu_\tau \overline{\nu_\ell}$, where $\ell = \mu, e$.  To better understand contributions of each set of new particles, we divide the numerical analysis to effects due to purely new gauge bosons, purely lepton flavor-changing scalar bosons, and purely lepton flavor-conserving scalar bosons.  We scan for the parameter space that can accommodate $\Delta a_\mu$ while being allowed by the constraint from lepton universality in the Michel decays of $\tau$.

The layout of this paper is as follows.  In Section~\ref{sec:model}, we introduce the model, discussing the extended $SU(2)_{\mu\tau}$ gauge sector, scalar sector, and lepton sector.  We show in Section~\ref{sec:muon-g-2} how various new particles can contribute to the muon $g - 2$ and explain the current discrepancy between experimental data and the SM expectation.  Section~\ref{sec:tau-michel-decay} discusses the constraint on the model from the Michel decays of $\tau \to \ell \nu_\tau \overline{\nu_\ell}$.  An extension of the model to explain neutrino mass and a discussion of collider phenomenology are given in Section~\ref{sec:neutrino-collider}.  Section~\ref{sec:summary} concludes with a summary of our findings in this work.  We collect detailed formulas of scalar field mass matrices in Appendix~\ref{sec:mass-matrices}, new gauge interactions of $\mu$ and $\tau$ in Appendix~\ref{sec:gauge-int}, and new Yukawa interactions of $\mu$ and $\tau$ in Appendix~\ref{sec:yukawa-int}.

\section{Model of Gauged $SU(2)_{\mu\tau}$ \label{sec:model}}

\begin{table}[tb]
\begin{tabular}{|c||c|c|c|c||c|c|c|}
\hline\hline
& \multicolumn{4}{c||}{SM leptons} & \multicolumn{3}{c|}{Scalars}
\\
\hline
& $L\!=\!L_{e}$ & $L^{\alpha}\!=\!\begin{bmatrix} L_{\mu} \\ L_{\tau} \end{bmatrix}$ & 
$e_{R}^{}$ & ${e_{R}^{}}^{\alpha}\!=\!\begin{bmatrix} \mu_{R}^{} \\ \tau_{R}^{} \end{bmatrix}$ 
& $\Phi_{0}$ & 
$\Phi^{\alpha}_{~\beta}$ 
& $S^{\alpha}$ 
\\ 
\hline \hline 
$SU(2)_{\mu\tau}$ & ${\mathbf 1}$ & ${\mathbf 2}$ & ${\mathbf 1}$ & ${\mathbf 2}$ 
& ${\mathbf 1}$ & ${\mathbf 3}$ & ${\mathbf 2}$ 
\\ 
\hline 
$SU(2)_L$ & ${\mathbf 2}$ & ${\mathbf 2}$ & ${\mathbf 1}$ & ${\mathbf 1}$ 
& ${\mathbf 2}$ & ${\mathbf 2}$ & ${\mathbf 1}$ 
\\ 
\hline 
$U(1)_Y$ & $-1/2$ & $-1/2$ & $-1$ & $-1$ & $+1/2$ & $+1/2$ & $0$ 
\\ 
\hline\hline 
\end{tabular}
\caption{The particle content of the lepton and scalar sectors of the model and their quantum numbers under the $SU(2)_{\mu\tau}$ and SM $SU(2)_L$ and $U(1)_Y$ gauge groups.  All these particles are colorless under $SU(3)_C$, and all the other SM particles are singlets under the $SU(2)_{{\mu\tau}}$.}
\label{Tab:MuTau}
\end{table}

In addition to the SM gauge group $SU(3)_C \times SU(2)_L \times U(1)_Y$, the model considered in this work has a gauged $SU(2)_{\mu\tau}$ symmetry, which is the lepton flavor symmetry within the second- and third-generation leptons. 
To break the $SU(2)_{\mu\tau}$ symmetry, we introduce a scalar field $S$, which is a singlet under the SM gauge group and a doublet under the gauge flavor symmetry.  
Moreover, to break the degeneracy in the mass of the second- and third-generation leptons, the model also has another new scalar field $\Phi$ that transforms as an $SU(2)_L$ doublet and an $SU(2)_{\mu\tau}$ triplet.
The new scalar fields are all colorless.  
Quantum numbers of the lepton and scalar fields under the $SU(2)_{{\mu\tau}}$, $SU(2)_L$ and $U(1)_Y$ groups are listed in Table.~\ref{Tab:MuTau}.  
Also explicitly shown are the $SU(2)_{{\mu\tau}}$ indices $\alpha,\beta \in \{ 1,2 \}$.  
In this paper, we reserve the first few Greek letters for the flavor indices of $SU(2)_{\mu\tau}$ and the others for the Lorentz indices.
We note in passing that with the above-mentioned particle content, the model is free of gauge anomalies. 
\footnote{The Witten anomaly is known to occur if an $SU(2)$ model contains only an odd number of doublets charged under it.  In our $SU(2)_{\mu\tau}$ model, this anomaly disappears when we introduce the flavor-doublet right-handed neutrinos, which are required to generate active neutrino mass by the seesaw mechanism, to be discussed in Section~\ref{sec:neutrino-collider}. }

The most general scalar potential of $\Phi_{0}, \Phi$ and $S$ is given by
\begin{align}
V(\Phi_{0}, \Phi, S) = 
&
-\mu_{S}^{2} |S|^{2} 
-\mu_{0}^{2} \Phi_{0i}^{\dag} \Phi_{0}^{i}
+2m_{\Phi}^{2}\, \text{Tr}(\Phi^{\dag}_{i}\Phi^{i}) 
+\lambda_{S}^{} |S|^{4} 
+ \lambda_{0S}^{} |S|^{2}\, \Phi_{0i}^{\dag}\Phi_{0}^{i} \nonumber \\
& 
+2 \lambda_{\Phi S1}^{} |S|^{2}\, \text{Tr}(\Phi^{\dag}_{i}\Phi^{i}) 
+2 \lambda_{\Phi S2}S^{\dag} [\Phi^{\dag}_{i}, \Phi^{i}] S 
\nonumber \\
& 
+2\lambda'_{1} \big(\text{Tr}(\Phi_{i}^{\dag}\Phi^{i}) \big)^{2} 
+2\lambda''_{1} \text{Tr}(\Phi_{i}^{\dag}\Phi^{j}) \text{Tr}(\Phi_{j}^{\dag}\Phi^{i})   
+2\lambda'''_{1} \text{Tr}(\Phi_{i}^{\dag}\Phi_{j}^{\dag}) \text{Tr}(\Phi^{i}\Phi^{j}) \nonumber \\
&  
+\frac{\lambda_{2}^{}}2 (\Phi_{0i}^{\dag} \Phi_{0}^{i})^{2} 
+2\lambda_{3}^{} \text{Tr}(\Phi_{j}^{\dag}\Phi^{j})\, \Phi_{0i}^{\dag}\Phi_{0}^{i}
+2\lambda_{4} \text{Tr}(\Phi^{i} \Phi_{j}^{\dag})\, \Phi_{0i}^{\dag}\Phi_{0}^{j}
\nonumber \\
&
+ \Big\{ 
\lambda'_{\Phi S}S\, i\,\sigma_{2} [\Phi^{\dag}_{i}, \Phi^{i}] S
+ 
\lambda_{5} \text{Tr}(\Phi_{i}^{\dag}\Phi_{j}^{\dag})\, \Phi_{0}^{i}\Phi_{0}^{j} 
+ 4\,\kappa\, \Phi_{0i}^{\dag} \text{Tr}(\Phi_{j}^{\dag}\Phi^{j}\Phi^{i}) 
\nonumber \\
& \qquad
+\lambda_{\mu} \Phi_{0i}^{\dag} S^{\dag} \Phi^{i} S
+ \lambda'_{\mu1} \Phi_{0i}^{\dag} S\,i\sigma_{2} \Phi^{i} S
+ \lambda'_{\mu2} S\,i\sigma_{2} \Phi_{i}^{\dag} S\, \Phi_{0}^{i} +\text{H.c.} \Big\}
~, 
\end{align}
where $\lambda'_{\Phi S}$, $\lambda_{5}$, $\kappa$, $\lambda_{\mu}$, $\lambda'_{\mu1}$ 
and $\lambda'_{\mu2}$ are generally complex parameters. 
Hereafter, we take all of them to be real for simplicity. 
The $SU(2)_{L}$ indices are shown explicitly by $i, j$. 
The flavor adjoint Higgs doublet $\Phi$ is parameterized as 
\begin{align}
\Phi^{\alpha}_{~\beta}{}
= \frac{(\sigma_{a}^{})^{\alpha}_{~\beta}}2\, \Phi_{a} 
= \frac12 
\begin{bmatrix} 
\Phi_{3} & \sqrt2 \Phi_{+} \\ 
\sqrt2\Phi_{-} & -\Phi_{3} 
\end{bmatrix}, 
\quad
\Phi_{\pm}=(\Phi_{1}\mp i\,\Phi_{2})/\sqrt2,
\end{align}
where $a=1,2,3$, and the $SU(2)_{\mu\tau}$ flavor space is denoted by the square brackets.  
We assume that the SM singlet field $S$ develops a VEV
at an energy higher than the electroweak scale.  
Without loss of generality, we parameterize the $S$ field as 
\begin{align}
S^{\alpha} 
= \begin{bmatrix} S_{+} \\ S_{0} \end{bmatrix}
= \frac1{\sqrt2} \begin{bmatrix} S_{+}^{h}+i\,S_{+}^{z} \\ v_{S}^{}+S_{0}^{h}+i\,S_{0}^{z} \end{bmatrix},
\end{align}
where $S_{-}\equiv(S_{+})^{\star}$, $S_{0}^{z}$ will be identified as a would-be 
Nambu-Goldstone (NG) boson corresponding to the longitudinal mode of the new flavor gauge boson $X_{3}^{\lambda}$, and the superscripts $h, z$ here denote respectively the real and imaginary components of the field. 
The VEV of $S$ breaks the $SU(2)_{\mu\tau}$ symmetry completely. 
The flavor-singlet Higgs doublet field $\Phi_0$ is assumed to develop a VEV 
at the electroweak scale, and can be parameterized as
\begin{align}
\Phi_{0} = \begin{pmatrix} \phi_{0}^{+} \\ \phi_{0}^{0} \end{pmatrix}
= 
\begin{pmatrix} 
i\, \omega_{0}^{+} \\ 
\displaystyle \frac{v_{0}^{} + h_{0}^{} + i\,z_{0}^{}}{\sqrt2} 
\end{pmatrix}
~.
\end{align}
The flavor adjoint Higgs doublet $\Phi$ is required to have VEV's as well in order to generate mass splitting between $\mu$ and $\tau$. 
In general, the VEV's of $\Phi$ can be induced by those of $S$ and $\Phi_0$.  
In particular, the VEV's of $\Phi_\pm$ will contribute to the off-diagonal elements of the charged lepton mass matrix and make the vacuum structure more complicated. 
For simplicity, we consider that only $\Phi_{3}$ is induced to have a VEV, $\langle \phi_3^0 \rangle = v_3 / \sqrt2$, as the easiest way to generate the mass splitting.  In this case, we have $v^{2}=v_{0}^{2}+v_{3}^{2} = (246~{\rm GeV})^2$ 
and $\tan\beta \equiv v_{0}^{}/v_{3}^{}$. 
To realize this simple setup, we set $\lambda'_{\Phi S}=\lambda'_{\mu1}=\lambda'_{\mu2}=0$ in the following discussions.~\footnote{
These terms can also be removed by imposing a $\mathbb{Z}_{3}$ charge ``$\omega$'' (or a new $U(1)$ charge) for the $S$ field.
}
We then parametrize the flavored Higgs and scalar doublet fields as 
\begin{align}
\Phi_{3} = \begin{pmatrix} \phi_{3}^{+} \\ \phi_{3}^{0} \end{pmatrix}
= 
\begin{pmatrix} 
i \,\omega_{3}^{+} \\ 
\displaystyle \frac{v_{3}^{}+h_{3}^{}-i\,z_{3}^{}}{\sqrt2} 
\end{pmatrix},
\quad
\Phi_{\pm} = \begin{pmatrix} \phi_{\pm}^{+} \\ \phi_{\pm}^{0} \end{pmatrix}, 
\end{align}
where we take the phase convention that $\phi^{-}_{\mp} \equiv (\phi^{+}_{\pm})^{*}$. 
Note that in the limit where $S$ does not develop a VEV, the flavor-diagonal VEV of $\Phi$ induced by $\langle \Phi_0 \rangle$ would also break $SU(2)_{\mu\tau}$ to $U(1)_{\mathbb{L}_{\mu}-\mathbb{L}_{\tau}}$.~\footnote{
The global ``muon number'' symmetry $\mathbb{L}_{\mu}$ is realized and rearranged 
since $\mathbb{L}_{\mu} \langle S\rangle = 0$ and 
$[\, \mathbb{L}_{\mu}, \langle \Phi \rangle\, ] =0$, 
while the global ``tau lepton number'' $\mathbb{L}_{\tau}$ is not conserved in the scalar sector
because $\mathbb{L}_{\tau} \langle S\rangle \ne 0$, where
\begin{align}
(\mathbb{L}_{\mu})^{\alpha}_{~\beta} 
\equiv \frac{\delta^{\alpha}_{~\beta}+(\sigma_{3})^{\alpha}_{~\beta}}2 
= \begin{bmatrix} 1 & 0 \\ 0 & 0  \end{bmatrix},
\quad
(\mathbb{L}_{\tau})^{\alpha}_{~\beta} 
\equiv \frac{\delta^{\alpha}_{~\beta}-(\sigma_{3})^{\alpha}_{~\beta}}2 
= \begin{bmatrix} 0 & 0 \\ 0 & 1  \end{bmatrix}. 
\end{align}
The eigenvalues are calculated as
\begin{align}
(\mathbb{L}_{\mu})^{\alpha}_{~\beta} S^{\beta}
= \begin{bmatrix} 1 & 0 \\ 0 & 0  \end{bmatrix} 
\begin{bmatrix} S_{+} \\ S_{0} \end{bmatrix}
=\begin{bmatrix} S_{+} \\ 0 \end{bmatrix}, 
\quad  
[\, \mathbb{L}_{\mu}, \Phi\,] ^{\alpha}_{~\beta} 
= \frac12 \big[ \begin{bmatrix} 1 & 0 \\ 0 & 0  \end{bmatrix}, 
\begin{bmatrix} 
\Phi_{3} \!&\! \sqrt2\Phi_{+} \\ 
\sqrt2\Phi_{-} \!\!&\!\! -\Phi_{3} 
\end{bmatrix} \big]
=
\frac12 \begin{bmatrix} 
0 \!\!&\!\! \sqrt2\Phi_{+} \\ 
-\sqrt2\Phi_{-} \!\!&\!\! 0 
\end{bmatrix}.
\end{align}
Therefore, the subscripts $\pm,0\,(3)$ denote the muon number $\pm1,0\,(0)$, respectively, in the broken phase.  
} 
The scalar bosons in this model are thus classified according to their
(i) muon numbers, (ii) CP parities, and (iii) electric charges.

The mass eigenstates are defined in terms of the following field rotations:
\begin{align}
\begin{pmatrix}S_{0}^{h}\\h_3\\h_0\end{pmatrix} 
&=\text{R}(\alpha_{1},\alpha_{2},\alpha_{3})
\begin{pmatrix}s\\H\\h\end{pmatrix},\quad
\begin{pmatrix}z_3\\z_0\end{pmatrix}
=\text{R}(\beta) \begin{pmatrix}z\\A\end{pmatrix},\quad
\begin{pmatrix}\omega_3^+\\\omega_0^+\end{pmatrix}
=\text{R}(\beta) \begin{pmatrix}\omega^+\\H^+\end{pmatrix}, 
\end{align}
where $\omega^{\pm}$ and $z$ are the would-be NG bosons 
corresponding to the longitudinal modes of the $Z_{\lambda}$ and $W^{\pm}_{\lambda}$ bosons, and 
\begin{align}
\begin{split}
\text{R}(-\theta_{1},\theta_{2},-\theta_{3}) 
&=
\begin{pmatrix}
c_{1}c_{2} & s_{1}c_{2} & s_{2} \\
-(c_{1}s_{2}s_{3}+s_{1}c_{3}) & c_{1}c_{3}-s_{1}s_{2}s_{3} & c_{2}s_{3} \\
-c_{1}s_{2}c_{3}+s_{1}s_{3} & -(c_{1}s_{3}+s_{1}s_{2}c_{3}) & c_{2}c_{3}
\end{pmatrix}, \\
\text{R}(\theta)
&=
\begin{pmatrix}
\cos\theta&-\sin\theta\\
\sin\theta&\cos\theta
\end{pmatrix}, 
\end{split}
\end{align}
with the notation $(c_i,s_i) \equiv (\cos\theta_i,\sin\theta_i)$ for $i = 1,2,3$.
Moreover, the flavor off-diagonal Higgs bosons are given by
\begin{align}
\begin{pmatrix} S_{+} \\ \phi_{+}^{0} \\ \phi_{-}^{0\star} \end{pmatrix}
&=
\begin{pmatrix} 
c_{\beta'} & -c_{\alpha'}s_{\beta'} & s_{\alpha'}s_{\beta'} 
\\ 
\displaystyle \frac{s_{\beta'}}{\sqrt2} 
& \displaystyle \frac{c_{\alpha'}c_{\beta'}-s_{\alpha'}}{\sqrt2} 
& \displaystyle -\frac{s_{\alpha'}c_{\beta'}+c_{\alpha'}}{\sqrt2} 
\\ 
\displaystyle \frac{s_{\beta'}}{\sqrt2} 
& \displaystyle \frac{c_{\alpha'}c_{\beta'}+s_{\alpha'}}{\sqrt2} 
& \displaystyle -\frac{s_{\alpha'}c_{\beta'}-c_{\alpha'}}{\sqrt2}
\end{pmatrix}
\begin{pmatrix} \omega_{+} \\ H_{+} \\ h_{+} \end{pmatrix}, 
\end{align}
where $\tan\beta'\equiv 2v_{3}/v_{S}^{}$. 
Details of the mass matrices are given in Appendix~\ref{sec:mass-matrices}. 
We denote the $SU(2)_{\mu\tau}$ gauge fields by 
\begin{align}
(X^{\lambda})^{\alpha}_{~\beta}
= \frac{(\sigma_{a}^{})^{\alpha}_{~\beta}}2\, X_{a}^{\lambda} 
= \frac12 \begin{bmatrix} X_{3}^{\lambda} & \sqrt2 X_{+}^{\lambda} \\ 
\sqrt2X_{-}^{\lambda} & -X_{3}^{\lambda} \end{bmatrix}, 
\quad
X_{\pm}^{\lambda}= \frac{1}{\sqrt2}(X_{1}^{\lambda}\mp i\,X_{2}^{\lambda}). 
\end{align}
Again the subscripts $\pm$ associated with the new gauge bosons do not represent their electric charges. 
In fact, all the new gauge bosons are electrically neutral.  
The gauge interactions of matter fields are encoded in their kinetic terms involving the covariant derivative
\begin{align}
(\mathcal{D}^\lambda)^{\alpha}_{~\beta}
&= \delta^{\alpha}_{~\beta}\,D^\lambda + i\, g_{X}^{} (X^\lambda)^{\alpha}_{~\beta}~,
\label{eq:covariant-derivative}
\end{align}
where $D^{\lambda}$ is the covariant derivative in the SM, and $g_X^{}$ denotes the associated gauge coupling strength. 
The masses of new gauge bosons are generated by the VEV's of $\Phi$ and $S$ as
\begin{align}
+\frac12 \frac{g_{X}^{2}v_{S}^{2}}{4} X_{3}^{\lambda}X_{3\lambda} 
+\frac{g_{X}^{2}(v_{S}^{2}+4v_{3}^{2})}{4} X_{+}^{\lambda}X_{-\lambda}.
\label{eq:Xmasses}
\end{align}
This shows that $X_\pm$ is slightly heavier than $X_3$.
But as long as $v_{3}^{}\ll v_{S}^{}$, as assumed here, we can safely neglect this small difference and consider the new gauge bosons virtually degenerate in mass $\sim M_{X}$.  No mass mixing occurs between $X_3$ and the SM $Z$ boson because only $\Phi$ couples to both $SU(2)_L$ and $SU(2)_{\mu\tau}$ gauge bosons and the VEV of $\Phi$ breaks $SU(2)_{\mu\tau}$ to $U(1)_{\mathbb{L}_{\mu}-\mathbb{L}_{\tau}}$ associated with $X_3$.  The latter point is also reflected in the fact that the mass of $X_3$ does not receive the contribution of $v_3$, as shown in Eq.~\eqref{eq:Xmasses}.  Therefore, the model does not have $Z$-$X_3$ mixing at tree level. 
The new gauge interactions for $\mu$ and $\tau$ are given in Appendix~\ref{sec:gauge-int}.

There are two kinds of $SU(2)_{\mu\tau}$-invariant Yukawa interactions for $\mu$ and $\tau$:
\begin{align}
\overline{L}_{\alpha} (y_{0}\,\Phi_{0}\, \delta^{\alpha}_{~\beta} +2\,y\,\Phi^{\alpha}_{~\beta}) e_{R}^{\beta}
+\text{H.c.} 
\end{align}
The explicit form of the Yukawa interactions for $\mu$ and $\tau$ in the scalar mass eigenbasis 
is given in Appendix~\ref{sec:yukawa-int}. It is observed that in the limit of $\sin(\beta - \alpha) \simeq 1$ and large $\tan\beta$, where $\alpha \equiv \alpha_3$, the SM-like Higgs couplings to $\mu$ and $\tau$ are SM-like.  We note in passing that the ${\Phi^\alpha}_\beta$ and $S^\alpha$ fields do not directly couple with quarks due to their nontrivial $SU(2)_{\mu\tau}$ quantum charges.  
For definiteness, we will make the assumption of $\alpha_{1}=\alpha_{2}=0$ for the mixing angles.  Therefore, in this nearly alignment limit, the couplings of the SM-like Higgs boson $h$ with the quarks and electron 
are SM-like, while the extra CP-even Higgs boson $H$ are suppressed by a factor of 
$\cos(\beta-\alpha) - \sin(\beta-\alpha)/\tan\beta$ in the large $\tan\beta$ case.

In summary, we are considering in this model the alignment limit of $\sin(\beta - \alpha_3) \simeq 1$ and large $\tan\beta$ in order to be consistent with the current Higgs data, and taking the mixing angles $\alpha_1 = \alpha_2 = 0$ for simplicity.  For concreteness, we assume mass degeneracy for heavy scalar bosons unless we explicitly vary some of them.  Deviations from the above are not further investigated as the muon $g-2$ can already be accommodated within this parameter space.  We note that electroweak precision tests and other constraints, particularly those that could arise from significant nonzero couplings between the scalars and quarks, could be relevant in the general case.

\section{Muon $g - 2$ \label{sec:muon-g-2}}
%
We are now ready to discuss new contributions to the muon anomalous magnetic moment in this model. 
From the interactions listed in Appendices~\ref{sec:gauge-int} and \ref{sec:yukawa-int}, 
we expect three kinds of potentially large contributions from: 
(i) new gauge bosons $X_{3}^{\lambda}$ and $X_{\pm}^{\lambda}$, 
(ii) flavor-changing scalar bosons $h_{\pm}$ and $H_{\pm}$, and 
(iii) flavor-conserving scalar bosons $H$ and $A$. 
To evaluate the contributions from the scalar bosons, we impose the relations 
$2M^{2}/v_{S}^{2}=\lambda_{\Phi S1}v^{2}/v_{0}^{2}=\lambda_{0S}v^{2}/v_{3}^{2}$ ({\it i.e.}, $\alpha_{1}=\alpha_{2}=0$) for simplicity. 
{\it Effects of new gauge bosons $X_{3}^{\lambda}$ and $X_{\pm}^{\lambda}$}:
Contributions from the new gauge bosons are calculated as 
\begin{align}
\Delta a_{\mu}(X_{3}) 
&= 
\frac1{8\pi^{2}} \left( \! \frac{g_{X}^{}}2 \! \right)^{2} \varepsilon^{\mu}_{X} 
\int_{0}^{1} dx \frac{2x^{2}(1-x)}{(1-x)(1-\varepsilon^{\mu}_{X}x)+\varepsilon^{\mu}_{X}x}, \\
\Delta a_{\mu}(X_{\pm}) 
&= 
\frac1{8\pi^{2}} \left( \! \frac{g_{X}^{}}{\sqrt2} \! \right)^{2} \varepsilon^{\mu}_{X} 
\int_{0}^{1} dx \frac{2x(1-x)(x-2(1-\rho))+x^{2}(1+\rho-x)(1-\rho)^{2} \varepsilon^{\mu}_{X} }{
(1-x)(1-\varepsilon^{\mu}_{X}x)+\varepsilon^{\tau}_{X} x}, 
\end{align}
where $\varepsilon^{\ell}_{B} \equiv M_{\ell}^{2}/M_{B}^{2}$ and $\rho \equiv M_{\tau}/M_{\mu} \simeq 16.82$. 
The $X_{3}$ contribution is identical to the one in the $U(1)_{\mathbb{L}_{\mu}-\mathbb{L}_{\tau}}$ model, and is approximately given by  
$\Delta a_{\mu}(X_{3}) \approx (g_{X}^{2}/48\pi^{2})\varepsilon^{\mu}_{X}$ for $\varepsilon^{\mu}_{X}\ll1$. 
The sign of this new contribution is favored by the observed muon $g-2$ anomaly. 
The $X_{\pm}$ contributions are $(6\rho-4) \simeq 96.9$ times larger than 
that of $X_{3}$ due to the chirality flipping effect in the same limit.\
\footnote{In Ref.~\cite{Altmannshofer:2016brv}, phenomenological consequences of a new gauge boson with some properties similar to our $X_{\pm}$ have been discussed.  However, the muon number is not conserved in their interactions.}
Thanks to the muon number conservation, no lepton flavor violation is generated by 
the new gauge bosons.

\begin{figure}[tbh]
\centering 
\includegraphics[width=7.5cm]{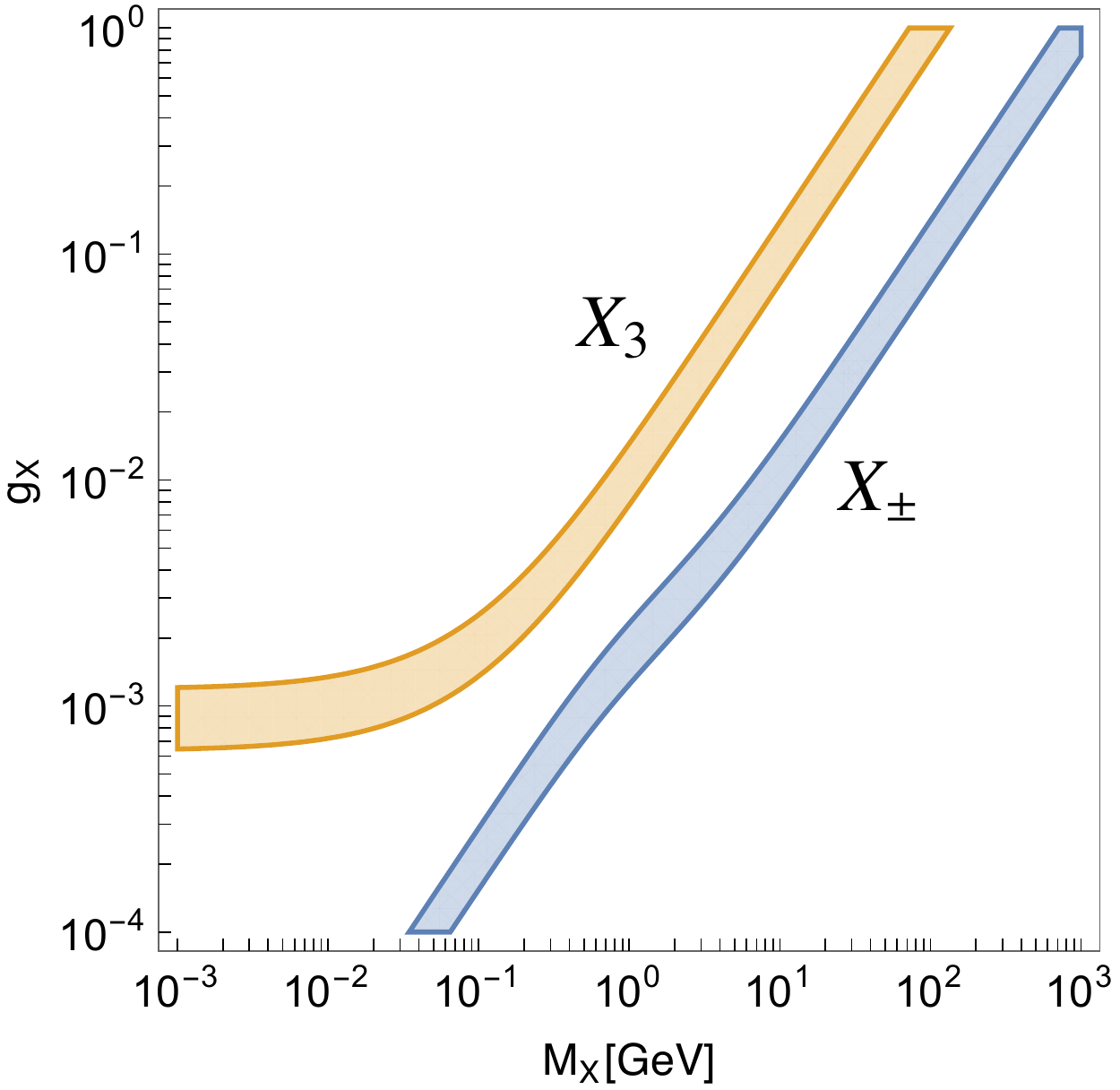}
\caption{Parameter space in the $(M_X,g_X)$ plane preferred by the muon $g-2$ anomaly at $2\sigma$ level.}
\label{FIG:Muon_g-2_X}
\end{figure}

Fig.~\ref{FIG:Muon_g-2_X} shows the regions in the $(M_X,g_X)$ plane that can accommodate the muon $g-2$ at the $2\sigma$ level ({\it i.e.}, $12.8<\Delta a_{\mu}^{\text{NP}}\times10^{10}<44.8$) with the contributions of the $X$ bosons. 
The orange (blue) region corresponds to the contribution from $X_{3}$ $(X_{\pm})$ alone. 
Since $\Delta a_{\mu}^{}(X_{\pm})\gg \Delta a_{\mu}^{}(X_{3})$, there is almost no change in the preferred parameter region from the blue region when contributions from all the $X$ bosons are taken into account.

{\it Effects of flavor-changing scalar bosons $h_{\pm}$ and $H_{\pm}$}:
The contributions from interactions with the flavor off-diagonal scalar bosons are listed as
\footnote{Effects of the flavor-changing {\it Higgs boson} in the type-III two-Higgs-doublet model have been studied in Ref.~\cite{Omura:2015nja}.  The {\it scalar} contributions discussed here is induced by $\Phi_{\pm}$ that does not develop a VEV.}
\begin{align}
\Delta a_{\mu}(h_{\pm})
&= 
\frac1{8\pi^{2}} \left( \frac{M_{\mu}(1-\rho)}{\sqrt2\, v\,c_{\beta}} \right)^{2} 
\varepsilon^{\mu}_{h_{\pm}^{}}
\int_{0}^{1}dx \frac{s_{\alpha'}^{2}c_{\beta'}^{2}(1-x+\rho) + c_{\alpha'}^{2}(1-x-\rho)
}{(1-x)(1-\varepsilon^{\mu}_{h_{\pm}^{}}x)+\varepsilon^{\tau}_{h_{\pm}^{}}x}x^{2}, \\
\Delta a_{\mu}(H_{\pm}) 
&= 
\frac1{8\pi^{2}} \left( \frac{M_{\mu}(1-\rho)}{\sqrt2\, v\,c_{\beta}} \right)^{2} 
\varepsilon^{\mu}_{H_{\pm}^{}}
\int_{0}^{1}dx \frac{c_{\alpha'}^{2}c_{\beta'}^{2}(1-x+\rho) + s_{\alpha'}^{2}(1-x-\rho)
}{(1-x)(1-\varepsilon^{\mu}_{H_{\pm}^{}}x)+\varepsilon^{\tau}_{H_{\pm}^{}}x}x^{2}. 
\end{align}
For $\varepsilon^{\ell}_{B}\ll1$ and $v\ll v_{S}^{}$, 
the sum of the two contributions is approximated by 
\begin{align}
\Delta a_{\mu\pm}
&\simeq 
\frac1{48\pi^{2}} \varepsilon^{\mu}_{h_{\pm}^{}} \!
\left( \frac{M_{\mu}(1-\rho)}{v} \sqrt{1+t_{\beta}^{2}}\, \right)^{2} \!
\nonumber \\
& \qquad\qquad
\times
\left\{ \! 1+R +3 \rho\, c_{2\alpha'}^{}
\left[ \! (1-R) \left( \! \ln \varepsilon^{\mu}_{h_{\pm}^{}}\!\! +\frac32 \right) \! -\! R \ln R
\right] \! 
\right\}, 
\label{Eq:a+-}
\end{align}
where $R=M_{h_{\pm}}^{2}/M_{H_{\pm}}^{2}$. 
The correction has an overall enhancement for large $\tan\beta$.
The third term in the curly brackets also receives a chirality flipping enhancement similar to 
the $X_{\pm}$ contribution.  Since this is proportional to $\rho$, the mass ratio of $m_\tau$ to $m_\mu$, 
it dominates in a wide region of parameter space.
Moreover, the sign of the contribution can be made positive by choosing an appropriate mixing angle $\alpha'$. 
The flavor-changing scalar interactions also respect the muon number conservation, 
so that lepton flavor violation is forbidden in this model.

\begin{figure}[tbh]
\centering
\includegraphics[width=5.3cm]{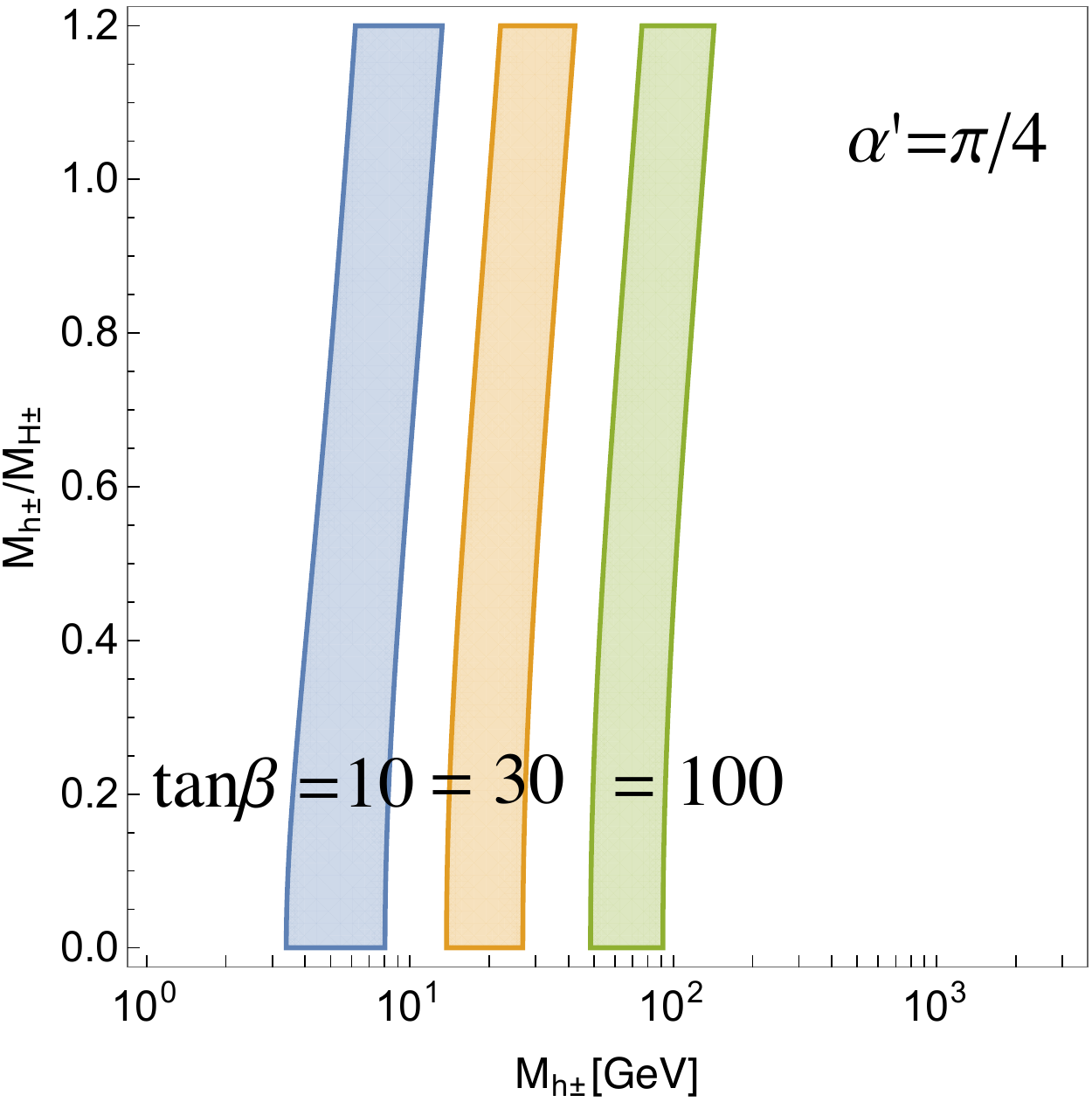} 
\includegraphics[width=5.3cm]{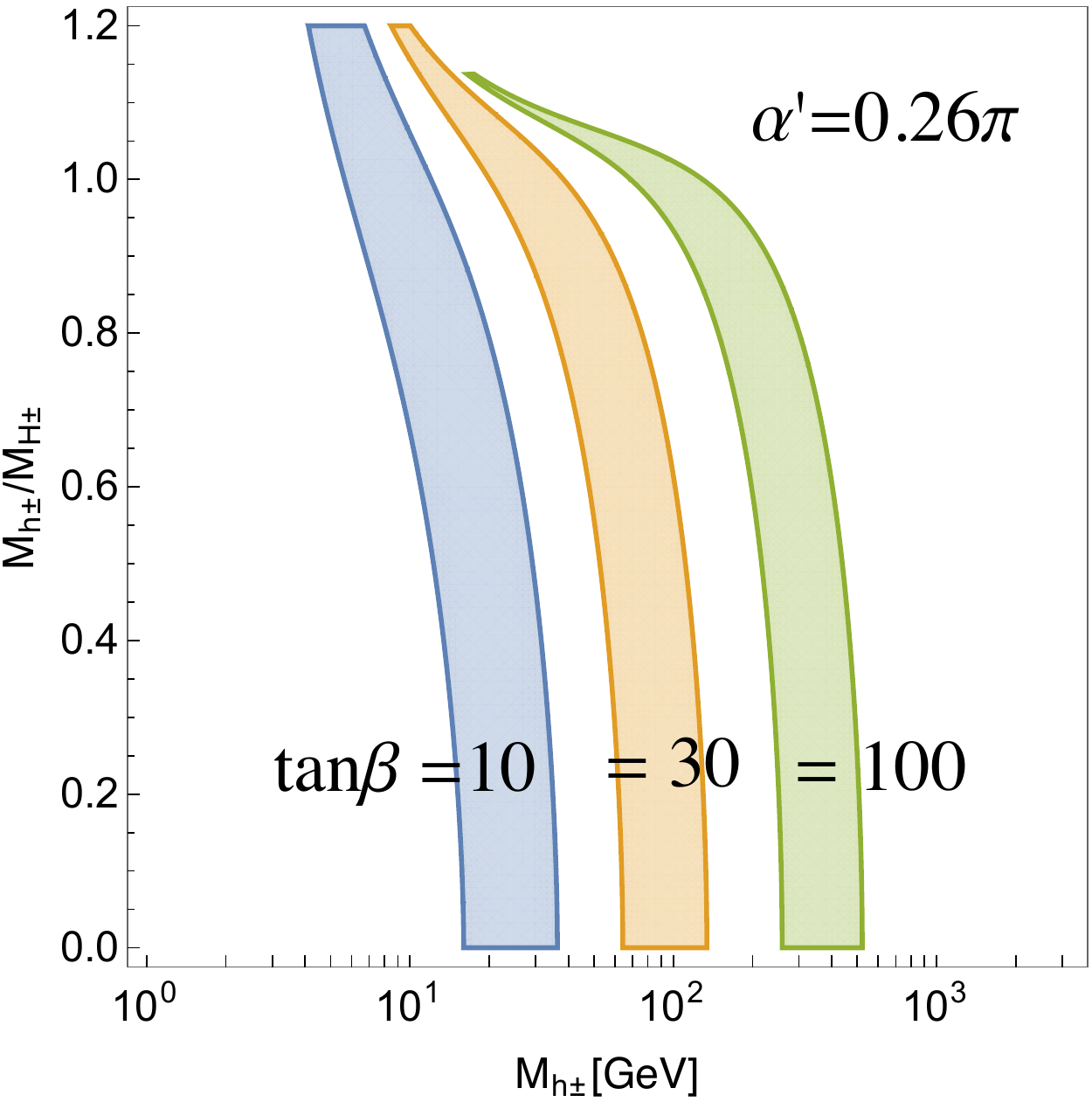}
\includegraphics[width=5.3cm]{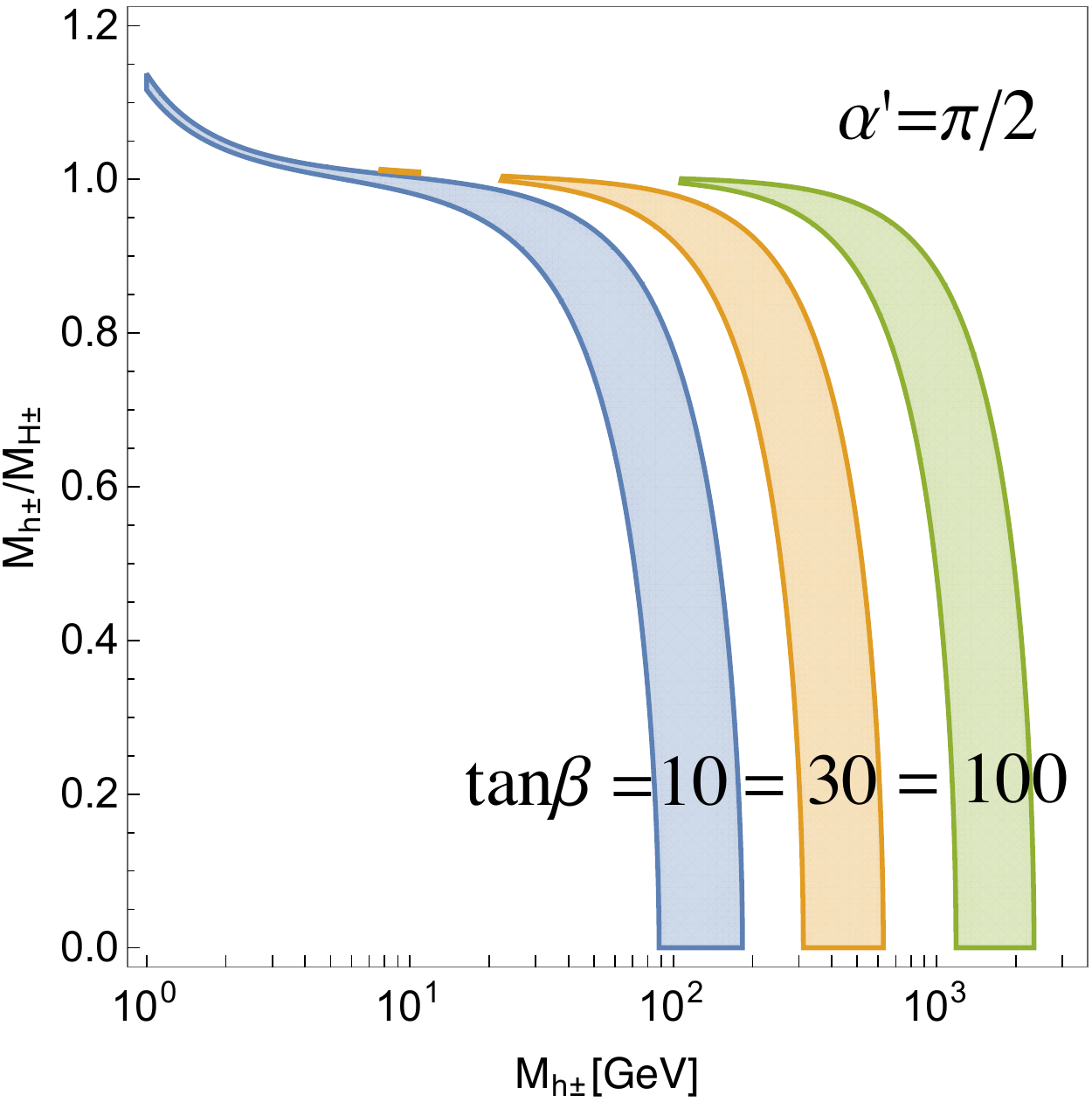} 
\\
\bigskip
\includegraphics[width=5.4cm]{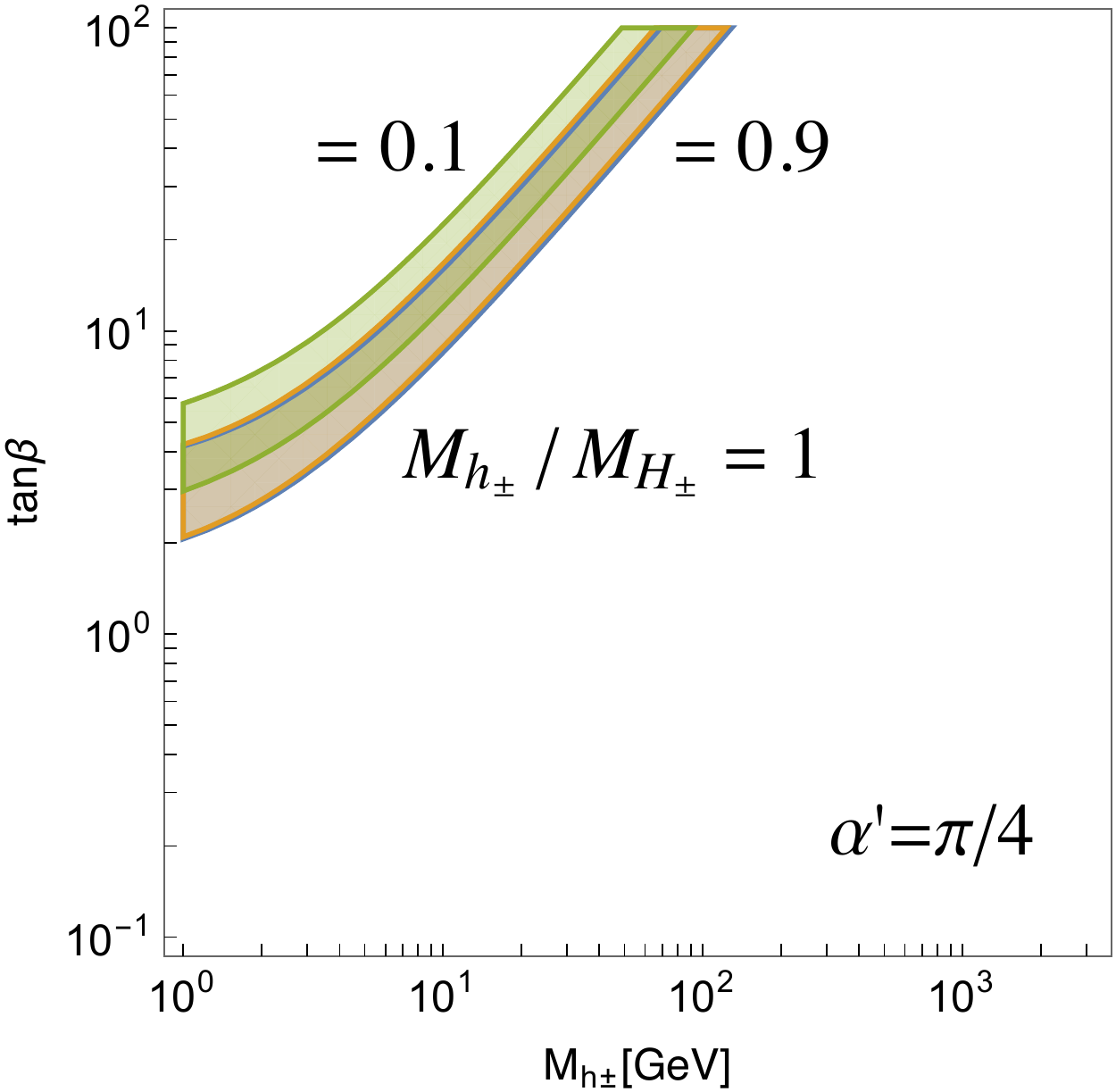}
\includegraphics[width=5.4cm]{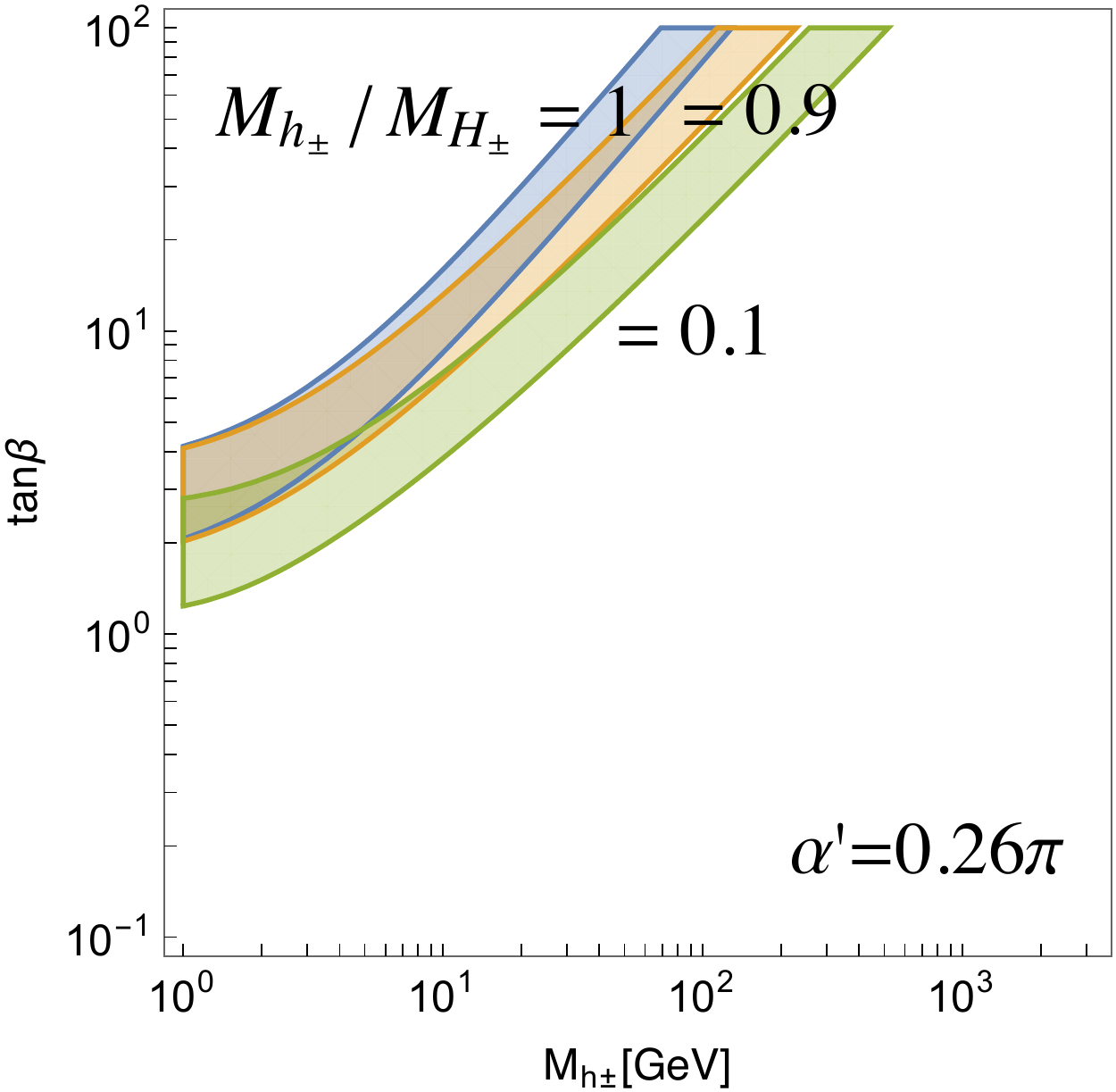} 
\includegraphics[width=5.4cm]{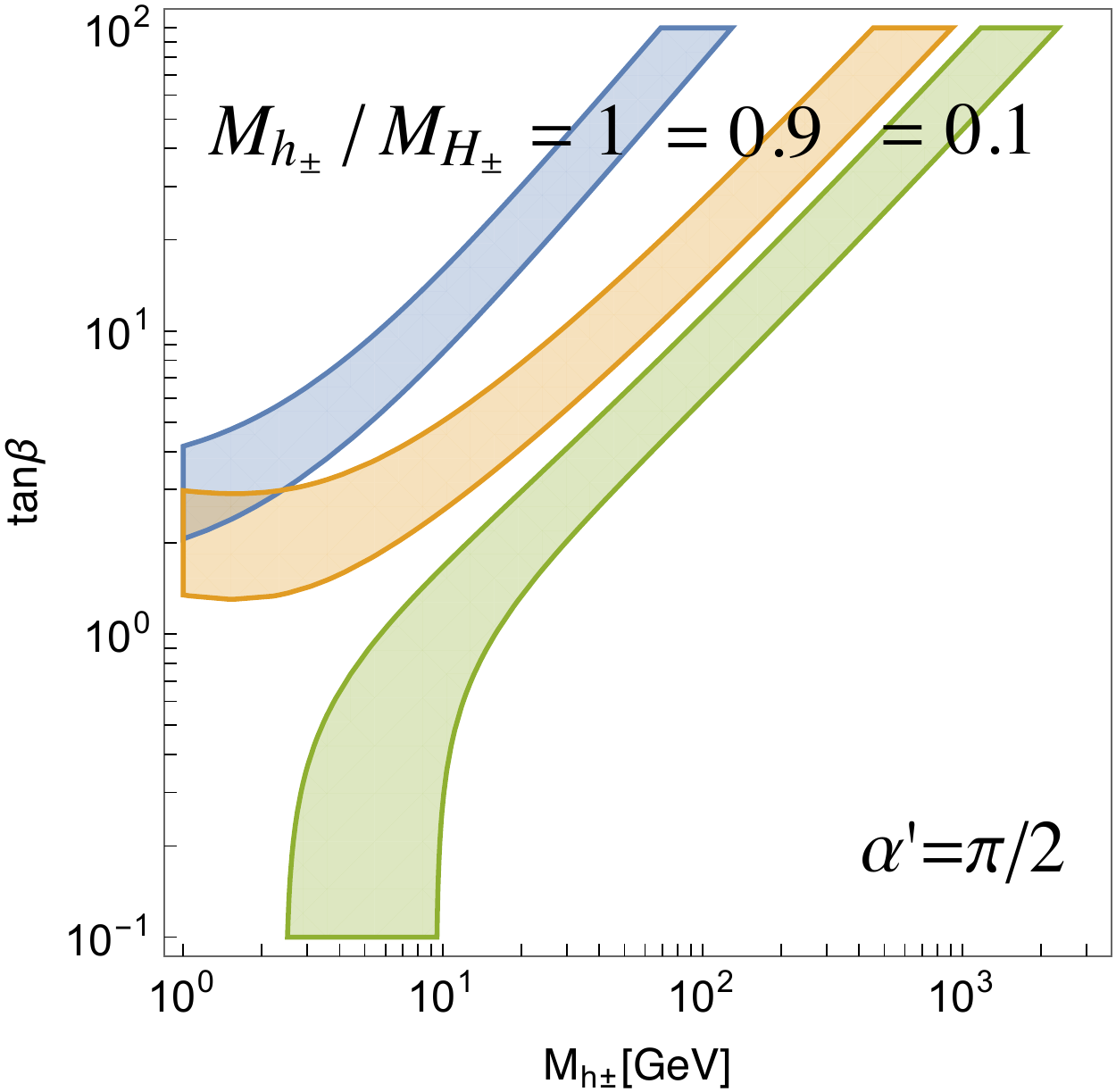} 
\caption{Upper row: parameter space in the $(M_{h_\pm},M_{h_\pm}/M_{H_\pm})$ plane for $\tan\beta = 10$ (blue), $30$ (orange) and $100$ (green), as preferred by the muon $g-2$ anomaly at $2\sigma$ level. Lower row: parameter space in the $(M_{h_\pm},\tan\beta)$ plane for the mass ratio $M_{h_\pm}/M_{H_\pm} = 1$ (blue), $0.9$ (orange) and 1 (green).  The left, middle, right plots are the cases with $\alpha'=\pi/4, 0.26\pi, \pi/2$, respectively. 
}
\label{FIG:Muon_g-2_phi}
\end{figure}

In Fig.~\ref{FIG:Muon_g-2_phi}, the contributions to the muon $g-2$ from the flavor-changing $h_{\pm}$ and $H_{\pm}$ bosons are evaluated, with all colored regions satisfying the muon $g-2$ data at $2 \sigma$ level.  For definiteness, we have taken $\beta'=0$ (corresponding to the large $v_{S}^{}$ limit) in numerical evaluations.  The left, middle, right plots are drawn for the cases of $\alpha'=\pi/4, 0.26\pi, \pi/2$, respectively.  Comparing the left and middle plots, we observe that the correction has a more sensitive dependence on $\alpha'$ around $\pi/4$.  In the upper row, the blue, orange, and green regions correspond respectively to the choices of $\tan\beta=10, 30, 100$.  For $\alpha \simeq \pi/4$, the mass of $h^\pm$ is preferred by data to fall in the regime of a few tens to a few hundreds of GeV and has less dependence on the mass ratio $R$ when it is smaller than 1.  For $\alpha' = \pi/2$ in contrast, $M_{h^\pm}$ is of ${\cal O}(0.1 - 1)$ TeV unless $R > 1$.  The peculiar behavior of the allowed regions around $R = 1$ in this case originates from the sign change of the last term in the curly brackets of Eq.~\eqref{Eq:a+-}.  This effect is less prominent in the other two plots because $c_{2\alpha'} \approx 0$.

In the lower row of Fig.~\ref{FIG:Muon_g-2_phi}, the blue, orange, and green regions correspond respectively to the choices of $M_{h_{\pm}}/M_{H_{\pm}}=1, 0.9, 0.1$. 
In the limit $R\to1$, where $\alpha'$ becomes undefined, the contributions with dependence on $\alpha'$ vanish.  
In the large $\tan\beta$ regime of interest to us, the contribution to the muon $g-2$ simply scales as $\tan^{2}\beta$. 

{\it Effects of flavor-conserving scalar bosons $H$ and $A$}:
Flavor-conserving interactions of $H$ and $A$ with the muon are also enhanced by 
$\rho \tan\beta$. 
Thus, contributions of $H$ and $A$ may be non-negligible as 
compared with those from flavor-changing scalar bosons, and are evaluated to be
\begin{align}
\Delta a_{\mu}(H)
&= 
\frac1{8\pi^{2}} \left( \! \frac{M_{\mu}}{v} \! \right)^{2} \varepsilon^{\mu}_{H} \!\!
\left[ \! \! \left( \!\! -c_{\beta-\alpha}^{} \!+ \! \frac{s_{\beta-\alpha}^{}}{t_{2\beta}^{}} \right)
+\rho \frac{s_{\beta-\alpha}^{}}{s_{2\beta}^{}} \right]^{2} \!
\int_{0}^{1}dx \frac{x^{2}(2-x)}{(1-x)(1-\varepsilon^{\mu}_{H}x)+\varepsilon^{\mu}_{H}x}, \\
\Delta a_{\mu}(A)
&= 
\frac1{8\pi^{2}} \left( \! \frac{M_{\mu}}{v} \! \right)^{2} \varepsilon^{\mu}_{A} 
\left( \frac1{t_{2\beta}^{}} +\rho \frac1{s_{2\beta}^{}} \right)^{2} \!
\int_{0}^{1}dx \frac{-x^{3}}{(1-x)(1-\varepsilon^{\mu}_{A}x)+\varepsilon^{\mu}_{A}x}. 
\end{align}
It is interesting to note that each of the integrals in the above two expressions is divergent when $\varepsilon_{H,A} \to 0$.  Nevertheless, they have opposite signs and result in destructive interference between the contributions from $H$ and $A$.
Therefore, most of the contributions cancel out in the alignment limit, $s_{\beta-\alpha}=1$, with mass degeneracy $M_{\varphi}\equiv M_{H}=M_{A}$:
\begin{align}
\Delta a_{\mu0}
&\simeq 
\frac1{48\pi^{2}} \varepsilon^{\mu}_{\varphi} \left( \frac{2M_{\mu}}{v} \right)^{2} 
\left( \frac1{t_{2\beta}^{}} +\rho \frac1{s_{2\beta}^{}} \right)^{2}, 
\qquad
(s_{\beta-\alpha}=1 \text{ and } M_{\varphi}\equiv M_{H}=M_{A}). 
\end{align}
In both large and small $\tan\beta$ limits, $1/t_{2\beta}$ and $1/s_{2_\beta}$ become large.  In the former case of interest to us, they interfere destructively. 
If the contribution from the chirality flipping term in Eq.~\eqref{Eq:a+-} is small, 
the effects of flavor-conserving scalar bosons can be comparable to those of 
flavor-changing scalar bosons. 
However, we will show in the next section that such parameter space is ruled out by the Michel decays.

\begin{figure}[tbh]
\centering 
\includegraphics[width=7.4cm]{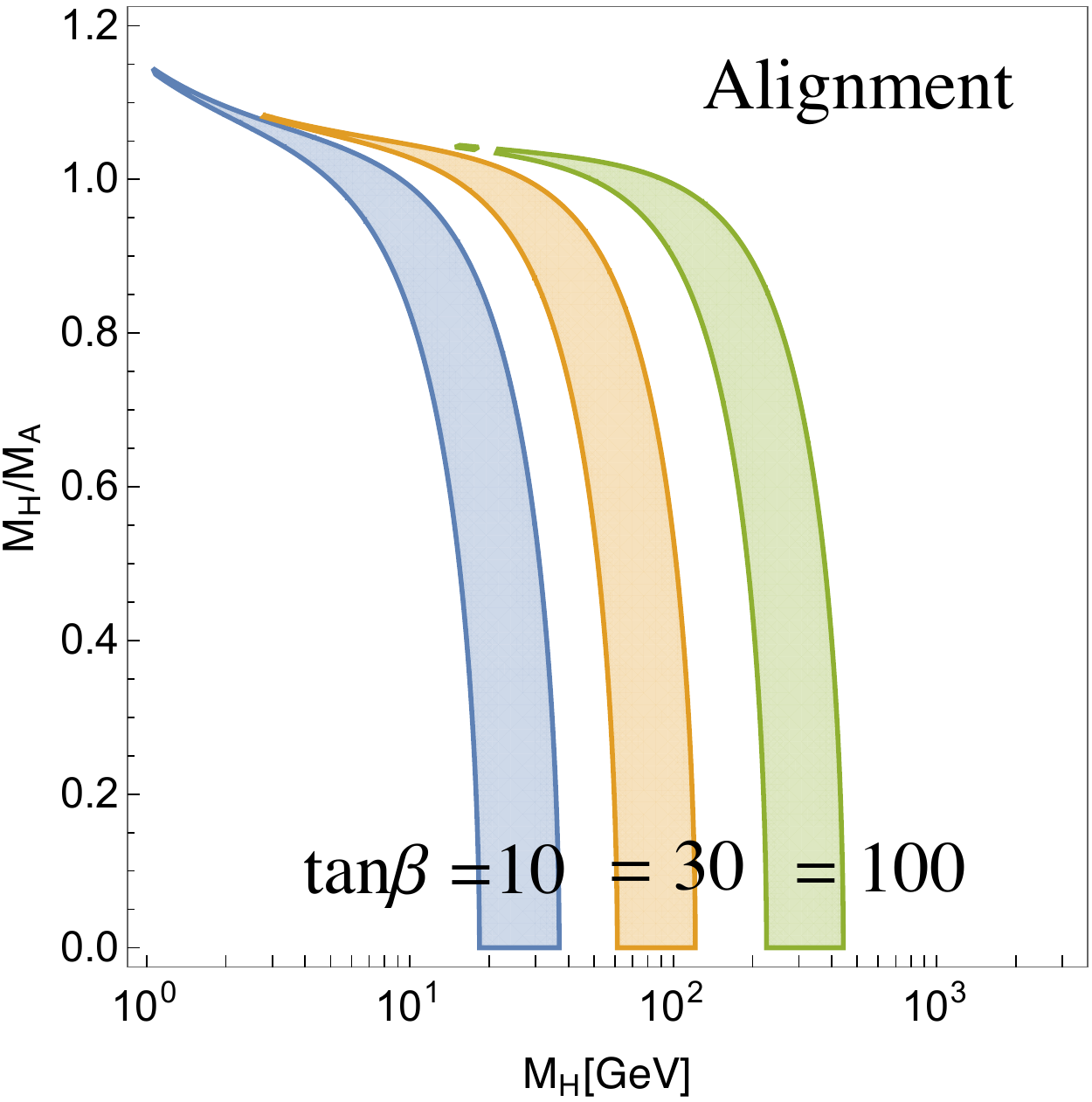}
\hspace{0.5cm}
\includegraphics[width=7.6cm]{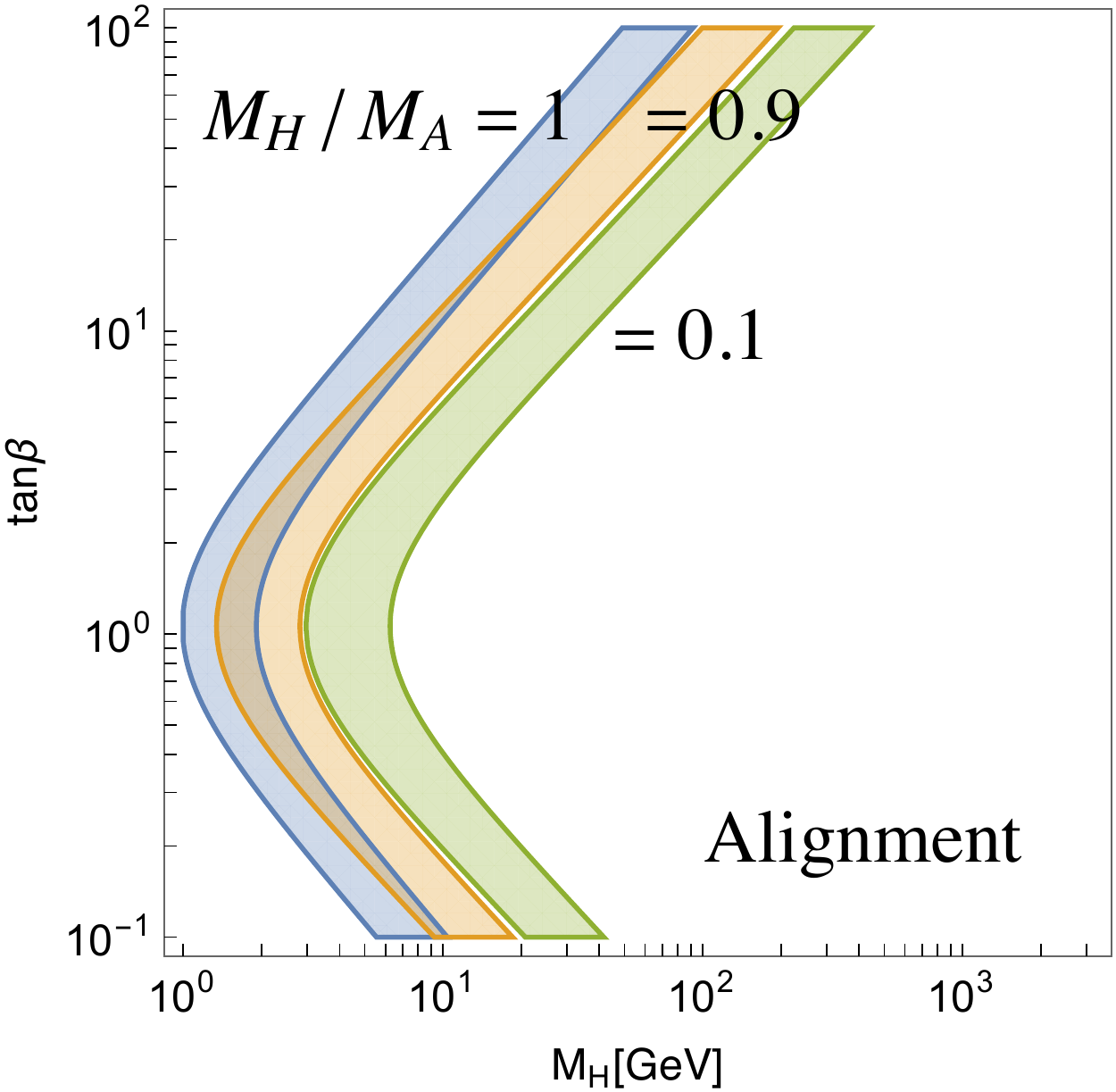}
\caption{Parameter space preferred by the muon $g-2$ anomaly at $2\sigma$ level.  Left: allowed regions in the $(M_H,M_A)$ plane for $\tan\beta = 10$ (blue), $30$ (orange) and 100 (green); right: allowed regions in the $(M_H,\tan\beta)$ plane for $M_H/M_A = 1$ (blue), $0.9$ (orange) and 0.1 (green).
}
\label{FIG:Muon_g-2_FlavorConservingScalar}
\end{figure}

In Fig.~\ref{FIG:Muon_g-2_FlavorConservingScalar}, we draw regions that can accommodate the muon $g-2$ anomaly at $2\sigma$ level.  The left plot shows a result similar to the analogous plot in Fig.~\ref{FIG:Muon_g-2_phi} for $\alpha' \simeq 0.26\pi$.  As noted above, $\Delta a_\mu$ gets an enhancement in either $\tan\beta \to 0$ or $\tan\beta \to \infty$ limit, thus the turning behavior in the allowed regions shown in the right plot.

\section{Michel Decays of $\tau$ \label{sec:tau-michel-decay}}
%
The Michel decays of the tau lepton can be modified by both $X_{\pm}$ and $H^{\pm}$ bosons at tree level. 
We consider the ratio
\begin{align}
R_{\tau\to\mu/\tau\to e} = 
\frac{\Gamma(\tau \to \mu \overline{\nu_{\mu}} \nu_{\tau})}
{\Gamma(\tau \to e \overline{\nu_{e}} \nu_{\tau})}
\end{align}
as the parameter showing violation in $\mu/e$ universality.  The measured value $R_{\tau\to\mu/\tau\to e}^{\rm exp} = 0.979 \pm 0.004$ is compared to the SM prediction of $0.9726$ with a comparatively negligible uncertainty~\cite{Patrignani:2016xqp}.

For definiteness, we consider two limiting scenarios: $M_{H^{\pm}} \gg M_{X}/g_{X}^{}$ and $M_{H^{\pm}} \ll M_{X}/g_{X}^{}$.  In the former case, the $\tau \to \mu \overline{\nu_{\mu}} \nu_{\tau}$ decay gets an additional contributing amplitude mediated by the $X_\pm$ gauge boson while the $\tau \to e \overline{\nu_{e}} \nu_{\tau}$ decay does not.  With the definition
\begin{align}
\frac{\alpha_X^{}}2 \equiv \frac{g_X^2}{4\sqrt2 G_F M_X^2} ~,
\end{align}
we find 
\begin{align}
\eta^X_{} \equiv 
\frac{R_{\tau\to\mu/\tau\to e}^{\rm X}}{R_{\tau\to\mu/\tau\to e}^\text{SM}}
=
\left( 1 + \alpha_X^{} + \frac12 \alpha_X^2 \right)
- 2\, \alpha_X^{} \frac1{\rho} \frac{g(\rho^{-2}_{})}{f(\rho^{-2}_{})}~,
\end{align}
where
\begin{align}
\begin{split}
f(z) 
&\equiv
1 - 8z + 8z^3 - z^4 - 12z^2\log z
~,
\\
g(z) 
&\equiv
1 + 9z - 9z^2 - z^3 + 6z \log z + 6 z^2 \log z
~. 
\end{split}
\end{align}
As numerically $\rho f(\rho^{-2}_{}) \gg g(\rho^{-2}_{})$, we observe constructive interference between the $W$ and $X_\pm$ contributions.  

\begin{figure}[tbh]
\centering 
\includegraphics[width=7.5cm]{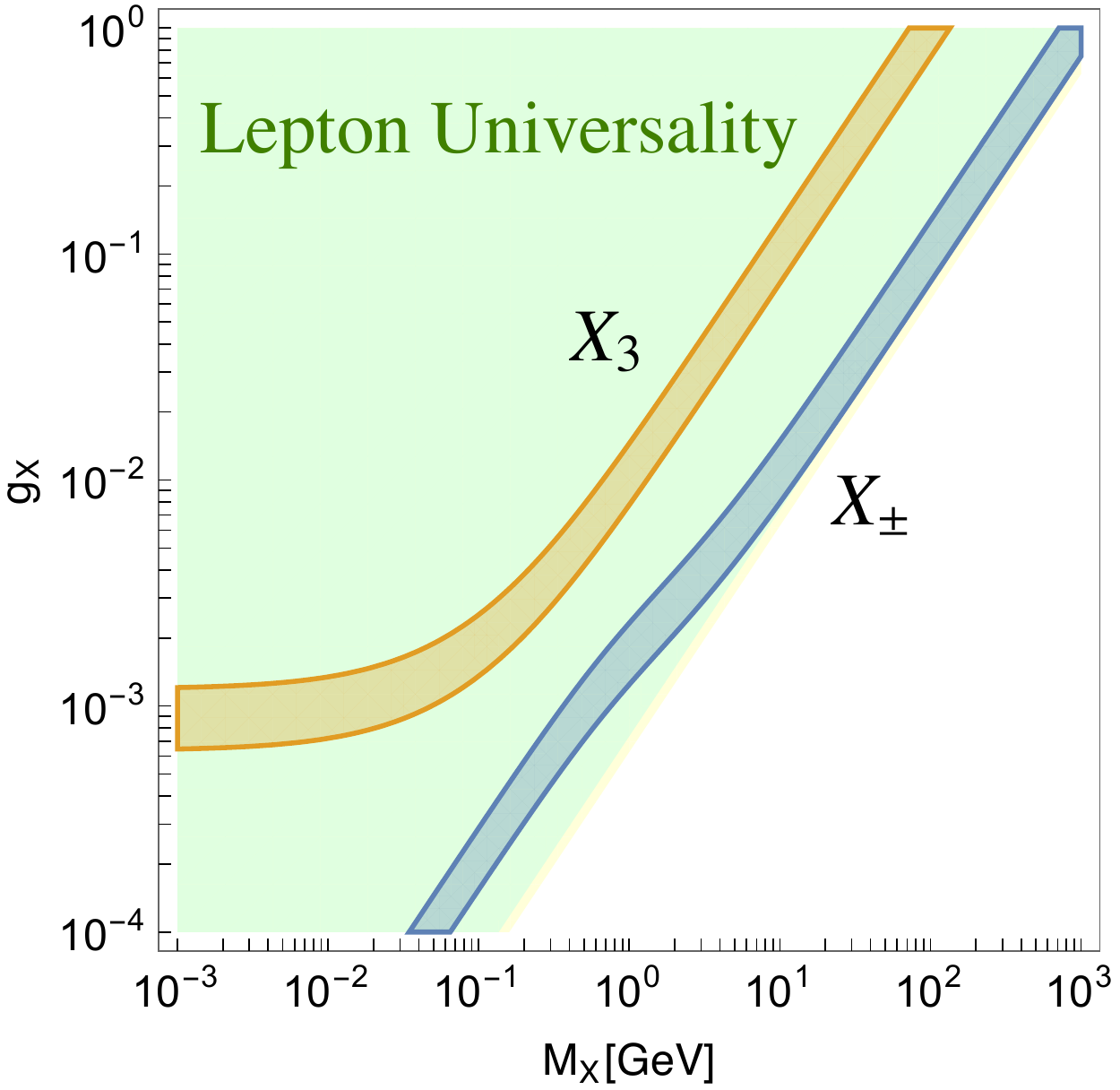}
\caption{Regions excluded by the constraint of $\mu/e$ lepton universality from the leptonic $\tau$ decays at the $1\sigma$ and $2\sigma$ levels are shown in light yellow and light green, respectively.  The regions favored by the muon $g-2$ anomaly are also superimposed.}
\label{FIG:LU_X}
\end{figure}

The experimental constraint of $\eta^{X} < 1.011~(1.015)$ at $1\sigma$ ($2\sigma$) level is shown in light yellow (light green) in Fig.~\ref{FIG:LU_X}.  As shown in the plot, the difference between the $1\sigma$ and $2\sigma$ bounds is very small.  It is seen that the entire parameter region favored by the muon $g-2$ anomaly is excluded.  Therefore, it is impossible to explain the muon $g-2$ anomaly while being consistent with the lepton universality in the Michel decays of $\tau$ with the $X$ gauge bosons alone.

In the limit of $M_{H^{\pm}} \ll M_{X}/g_{X}^{}$, we add only the contribution of $H^\pm$ to the SM amplitude and obtain
\begin{align}
\eta^{H^\pm}_{}
\approx& 
\Bigg\{ 
1
+\frac14 \left[ 
- \varepsilon_{H^{\pm}}^{\mu} 
\left( \frac{1-t_{\beta}^{2}}{2t_{\beta}} +\rho \frac{1+t_{\beta}^{2}}{2t_{\beta}} \right) 
\left( \rho \frac{1-t_{\beta}^{2}}{2t_{\beta}} +\frac{1+t_{\beta}^{2}}{2t_{\beta}} \right)
\right]^{2} \nonumber \\
&\qquad
+2
\left[ 
- \varepsilon_{H^{\pm}}^{\mu} 
\left( \frac{1-t_{\beta}^{2}}{2t_{\beta}} +\rho \frac{1+t_{\beta}^{2}}{2t_{\beta}} \right) 
\left( \rho \frac{1-t_{\beta}^{2}}{2t_{\beta}} +\frac{1+t_{\beta}^{2}}{2t_{\beta}} \right)
\right] \frac1{\rho} \frac{g(\rho^{-2})}{f(\rho^{-2})}
\Bigg\}, 
\end{align}
where we take the $M_{e}\to0$ limit. 
Note that the interference term is further suppressed by $1/\rho$.  Thus, the bound from the $\tau$ Michel decays is expected to be weak.

\begin{figure}[tbh]
\centering 
\includegraphics[width=5.4cm]{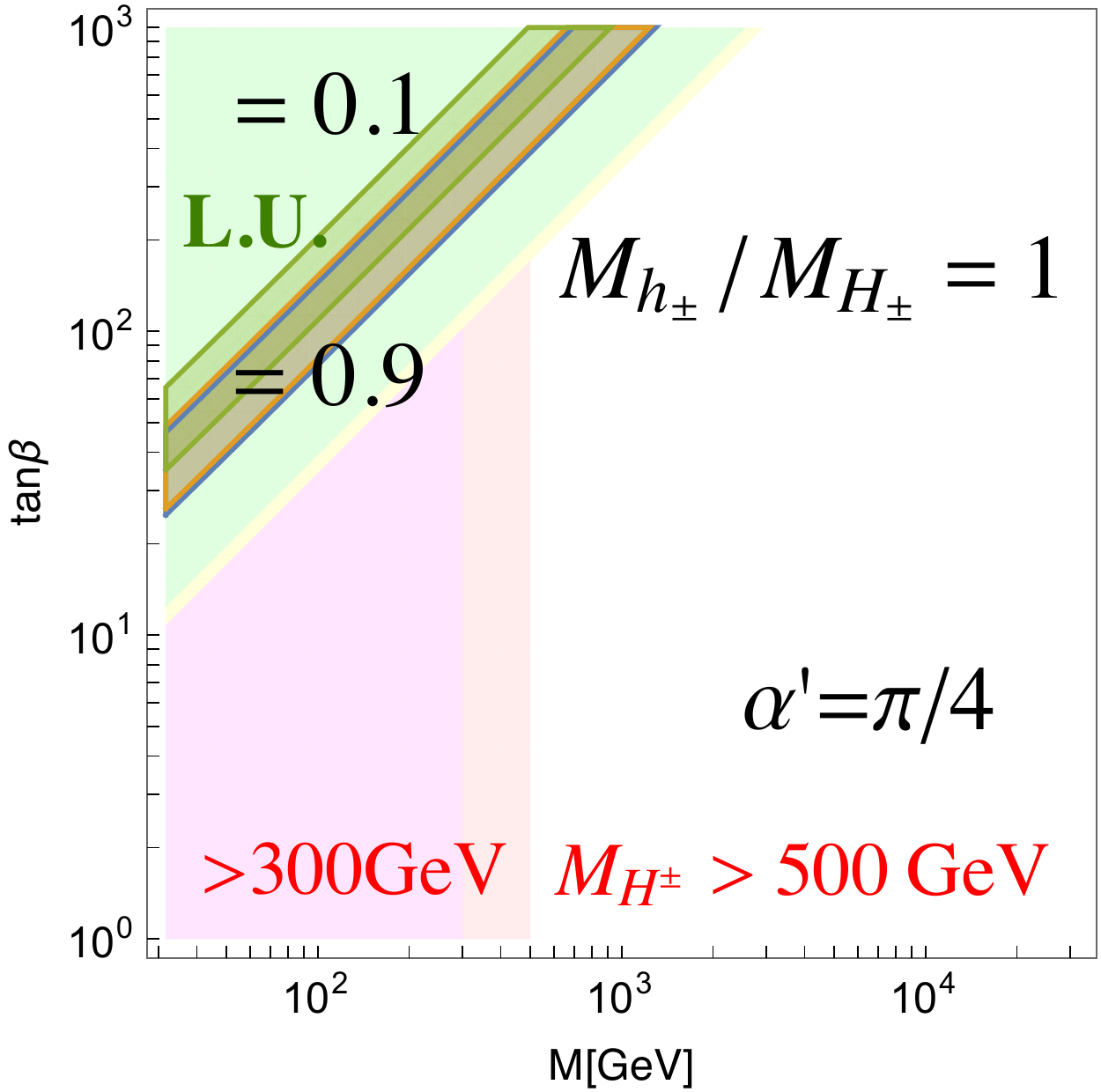}
\includegraphics[width=5.4cm]{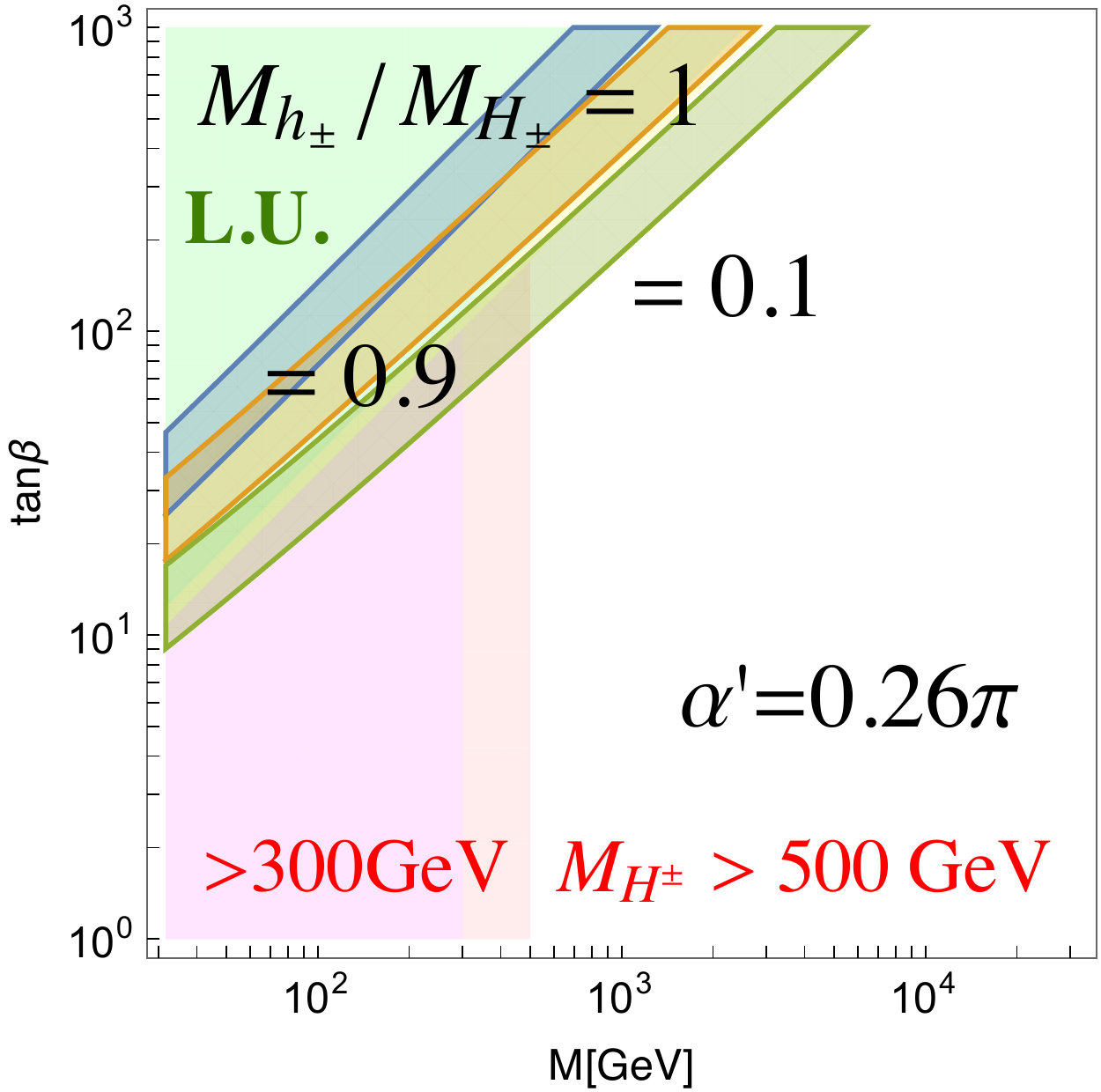} 
\includegraphics[width=5.4cm]{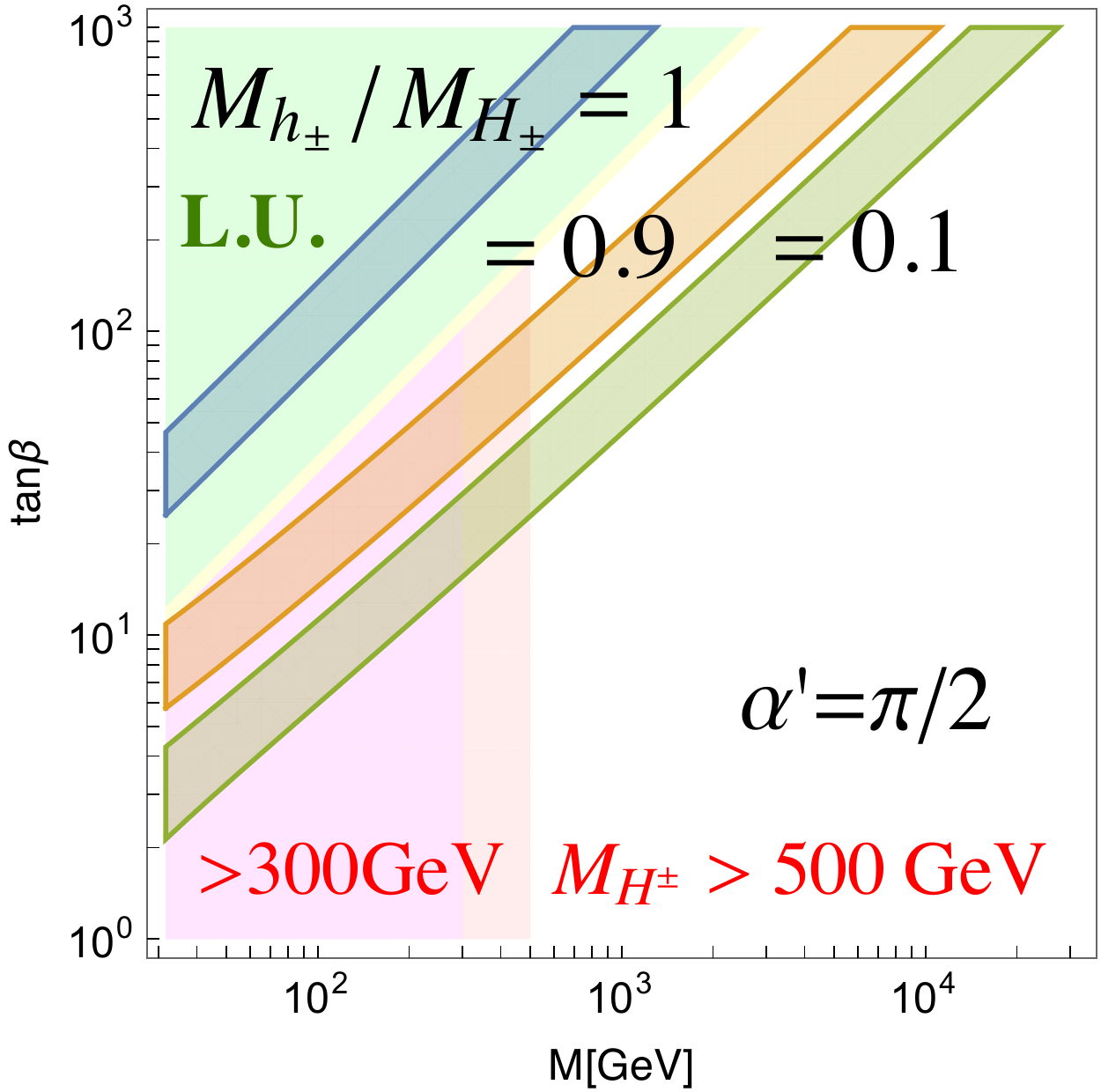} 
\caption{Parameter space favored by the muon $g-2$ anomaly at $2\sigma$ level is plotted together with the constraint of the $\mu/e$ lepton universality from the leptonic $\tau$ decays at $1\sigma$ (light green) and $2\sigma$ (light yellow) levels (labeled by L.U.) as well as the LHC direct search in light magenta for $M_{H^\pm} > 300$~GeV and light pink for $M_{H^\pm} > 500$~GeV.  
We here take into account only the contribution of flavor-changing scalar bosons to $\Delta a_\mu$, 
and assume $M^{2} \equiv M_{H}^{2}=M_{A}^{2}=M_{H^{\pm}}^{2}=M_{h_{\pm}}^{2}$ for definiteness.  
The left, middle, right plots are for $\alpha' = \pi/4, 0.26\pi, \pi/2$, respectively.
}
\label{FIG:LU_phi}
\end{figure}

The constraint of $\eta^{H^{\pm}}_{} < 1.011~(1.015)$ at $1\sigma$ ($2\sigma$) level along with the $\Delta a_\mu$ favored regions are drawn in Fig.~\ref{FIG:LU_phi} for the cases of $\alpha' = \pi/4$ (left plot), $0.26\pi$ (middle plot), and $\pi/2$ (right plot).
For definiteness, here we assume the mass relations $M^{2} \equiv M_{H}^{2}=M_{A}^{2}=M_{H^{\pm}}^{2}=M_{h_{\pm}}^{2}$.\footnote{
The parameter choice, $M_{H}^{2}\sim M_{A}^{2} \sim M_{H^{\pm}}$, is naturally realized 
in the decoupling limit, where the new mass scale in the Higgs sector controls their masses. 
This choice is also favored by the electroweak precision data, where their contributions to 
$S$ and $T$ parameters vanish in the degenerate mass limit~\cite{Kanemura:2011sj}.  
}
Only the contributions of flavor-changing scalar bosons to muon $g-2$ are taken into account 
because of the additional chirality enhancement and logarithmic enhancement seen in Eq.~\eqref{Eq:a+-}. 
We have also shown the direct search constraint on $H^{\pm}$, 
which would be expected to be the same as that from the slepton search from the electroweak production 
in the massless neutralino limit. 
We show here the left-handed selectron and smuon search lower mass bound of $300$~GeV 
in Ref.~\cite{Aad:2014vma} and the latest slepton mass lower bound of $500$~GeV 
(this bound would be an overestimate for our purpose since both left- and right-handed sleptons of all three flavors are summed over)~\cite{ATLAS:2017uun}. 
Such a direct search bound can be relaxed if the $H^{\pm} \to HW^{\pm}, AW^{\pm}$ decay modes are allowed when sufficient mass splitting exists among the particles.

\begin{figure}[tbh]
\centering 
\includegraphics[width=5.4cm]{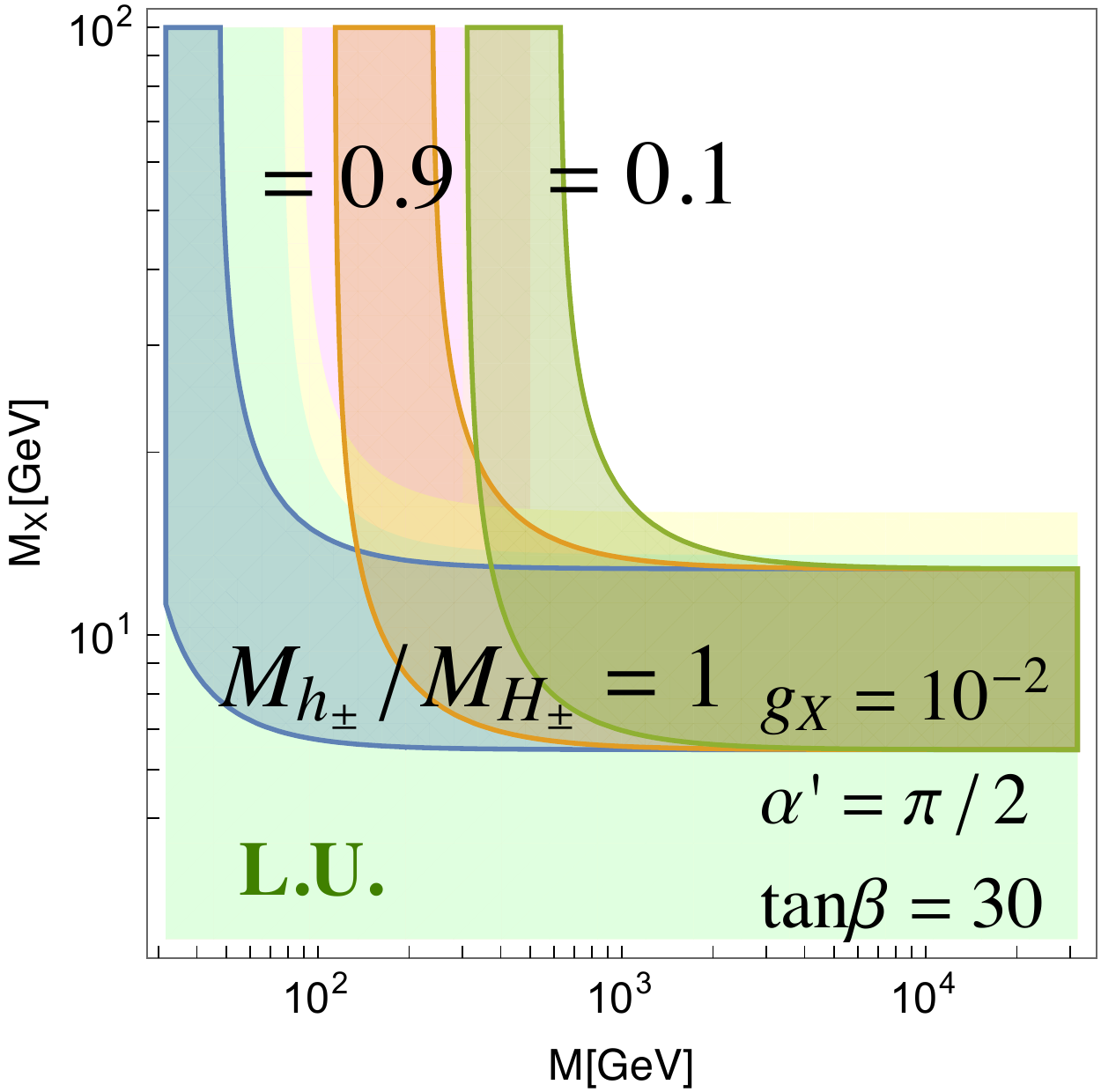}
\includegraphics[width=5.4cm]{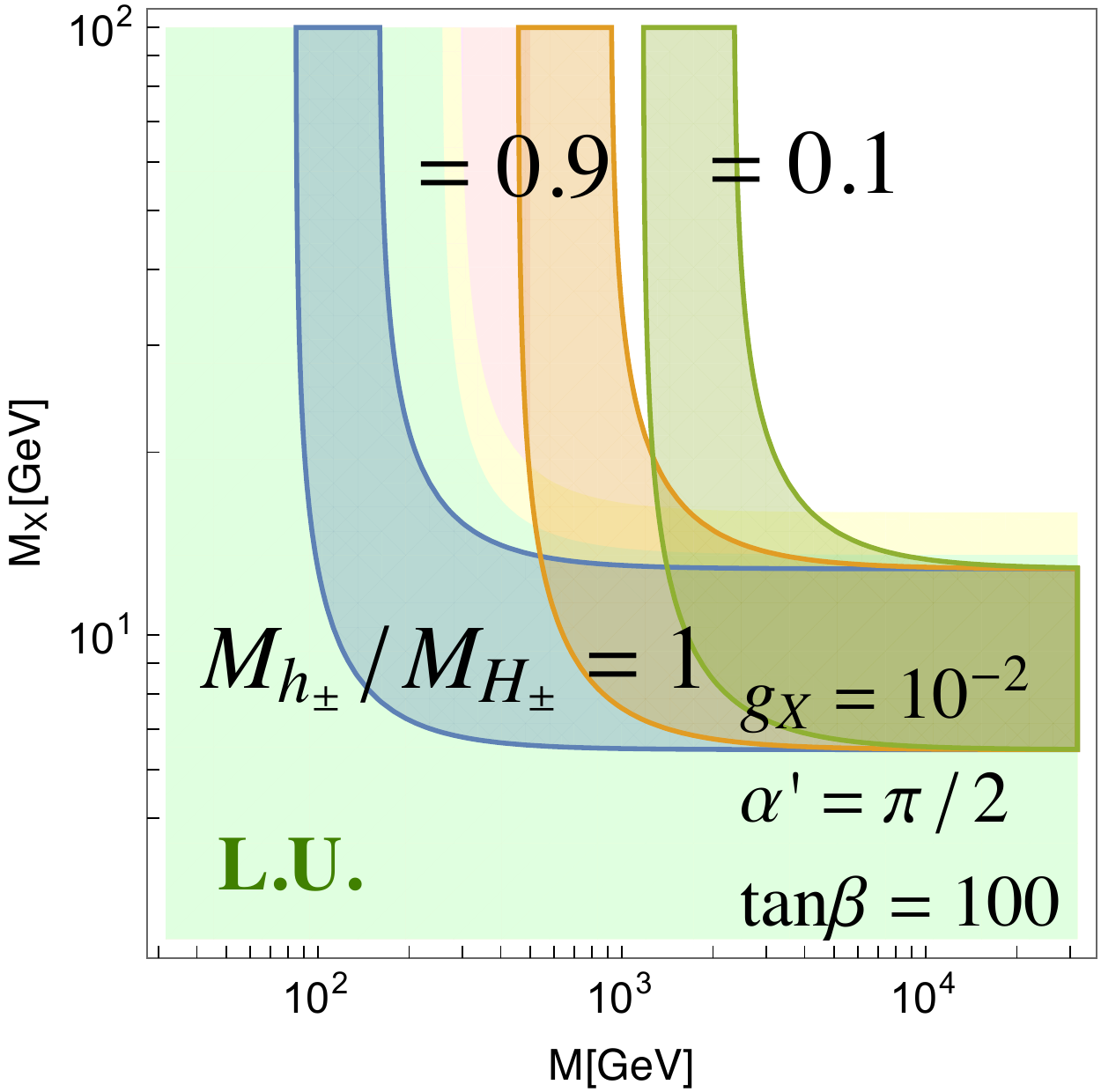} 
\includegraphics[width=5.4cm]{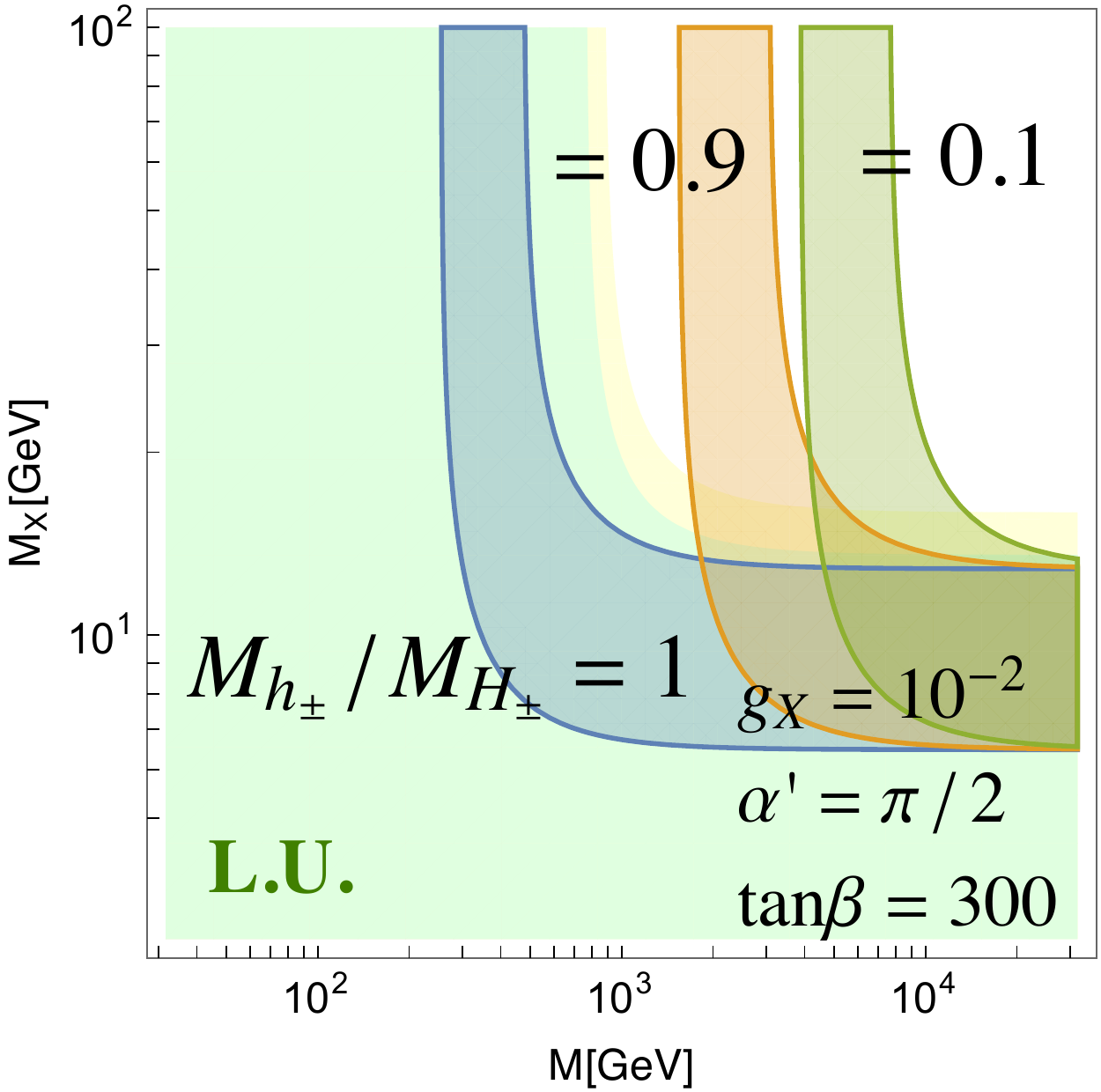} 
\caption{
Parameter space favored by $\Delta a_\mu$ for $M_{h_\pm} / M_{H_\pm} = 1$ (blue), $0.9$ (orange) and $0.1$ (green) and disfavored by the universality in leptonic $\tau$ decays at $1\sigma$ (light green) and $2\sigma$ (light yellow) levels.  The $M_{\pm} > 500$~GeV bound is drawn in light pink.  These plots all assume $g_{X}^{}=10^{-2}, \alpha'=\pi/2$ and $M^{2} \equiv M_{H}^{2}=M_{A}^{2}=M_{H^{\pm}}^{2}=M_{h_{\pm}}^{2}$ for definiteness.  The left, middle, right plots are for $\tan\beta = 30, 100, 300$, respectively. }
\label{FIG:LU_tanb}
\end{figure}

We note in passing that if $M_X/g_X$ and $M$ are comparable and therefore all $X_\pm$, $h_{\pm}$, and $H_{\pm}$ can contribute to the muon $g-2$ and the Michel decays of $\tau$ leptons, then it is possible to find some more allowed parameter space in the $(M_X,g_X)$ plane and the $(M,\tan\beta)$ plane. 
To see this cooperative effect clearly, we make plots in the $(M, M_{X})$ plane for fixed $(g_{X}^{}, \alpha')$ in Fig.~\ref{FIG:LU_tanb} for several values of $\tan\beta$.  In these plots, all the contributions from the new gauge bosons and scalar bosons are included in the calculations.  The choice of $g_{X}^{}=10^{-2}$ and $\alpha' = \pi/2$ is taken as an example.  A change in the value of $g_{X}^{}$ would only result in a different range of $M_{X}$.  We set the mixing angle $\alpha'=\pi/2$ in oder to show the maximal contribution to the muon $g-2$.  By varying $\alpha'$, one would obtain a smaller contribution.  From Fig.~\ref{FIG:LU_tanb}, we see that the contribution from the flavor-changing scalar is essential, thanks to the chirality flip and logarithmic enhancements.  The logarithmic enhancement reflects in the preference of a larger mass hierarchy between $H_\pm$ and $h_\pm$.  A comparison among the three plots indicates that $M_X$ needs to be greater than a few tens of GeV and that more parameter space is allowed for larger $\tan\beta$. 

{
\section{Neutrino and Collider Phenomenology\label{sec:neutrino-collider}}

%
The neutrino mass in the model can be generated through the usual Type-I seesaw mechanism.
The relevant terms for neutrino mass are given by
\begin{align}
-{\mathcal L}_{\nu}
&=
+\overline{L}\,\lambda_{e} \widetilde{\Phi_{0}} N_{R}
+\overline{L}_{\alpha}(\lambda_{+} \widetilde{\Phi_{0}} \delta^{\alpha}_{~\beta}
+2\lambda_{-} \widetilde{\Phi}^{\alpha}_{~\beta}) N_{R}^{\beta}
+\frac12M_{ee} \overline{N_{R}^{c}}N_{R} \nonumber \\
& \qquad
+\lambda_{e\mu} S^{\alpha}(i\sigma_{2})_{\alpha\beta}
\overline{N_{R}^{c}} N_{R}^{\beta}
+\lambda_{e\tau} S^{*}_{\alpha} \overline{N_{R}^{c}} N_{R}^{\alpha}
+\lambda_{\mu\tau} \overline{N_{R}^{c}}^{\alpha}(i\,\sigma_{2})_{\alpha\beta}
\Sigma^{\beta}_{~\gamma} N_{R}^{\gamma}
+\text{H.c.},
\end{align}
where $\widetilde{\Phi_0} \equiv i \sigma_2 \Phi_0^*$, 
$N_{R}$ and $N_R^\alpha$ are the right-handed $SU(2)_{\mu\tau}$ singlet and doublet neutrinos, respectively. Here we introduce another scalar field $\Sigma^{\beta}_{~\gamma}$ which is 
a SM singlet and $SU(2)_{\mu\tau}$ triplet that develops a VEV, $v_\Sigma$, as induced by that of $S$.\
\footnote{Here the diagonal VEV of $\Sigma$ is assumed to be induced in a way similar to that of $\Phi$, as can be arranged with a similar setup in the scalar potential.}
The Dirac mass and the Majorana mass matrices are given respectively by
\begin{align}
M_{D}= \frac1{\sqrt2}
\begin{pmatrix} \lambda_{e} v_{0} & 0 & 0 \\
0 & \lambda_{+}v_{0}+\lambda_{-}v_{3} & 0 \\
0 & 0 & \lambda_{+}v_{0}-\lambda_{-}v_{3} \end{pmatrix},
~~~
M_{N}=
\begin{pmatrix} 
M_{ee} & -\lambda_{e\mu} v_{S}^{} & \lambda_{e\tau} v_{S}^{} \\
-\lambda_{e\mu} v_{S}^{} & 0 & \lambda_{\mu\tau} v_{\Sigma}^{} \\
\lambda_{e\tau} v_{S}^{} & \lambda_{\mu\tau} v_{\Sigma}^{} & 0 
\end{pmatrix}.
\end{align}
The structures of these mass matrices are exactly the same as those discussed
in Ref.~\cite{Asai:2017ryy}.
Note that the ``muon number'' of $N_{R}$ can be defined through the $\lambda_{e\mu}$ coupling.  However, the muon number is softly broken by the $M_{ee}$ term.  With such an extension in the neutrino sector, the muon number is no longer a conserved charge.

Here we make some comments about collider searches of the new gauge bosons and scalar bosons in the model.  
Since the $X_{3,\pm}$ gauge bosons couple to $\mu$, $\tau$ and neutrinos, 
they can be produced at the CERN LHC in association with Drell-Yan $\mu, \tau$ or $\nu$ pairs 
from final-state bremsstrahlung, followed by leptonic decays and leading to  multi-lepton final states (without/with missing energy)~\cite{Harigaya:2013twa}. 
The relation of the lepton invariant mass $M_{\mu\mu}\sim M_{\tau\tau} \sim M_{X}$ would be a first step to confirm the diagonal subgroup of $SU(2)_{\mu\tau}$. 
In addition, our model predicts a peculiar decay channel $X_{\pm} \to \mu \tau$ with $M_{\mu\tau}\sim M_{X}$. 
This signal is the smoking-gun signature of the proposed model.\
\footnote{
A similar process for flavon has been discussed in the literature~\cite{Muramatsu:2017xmn}. 
}
The beam dump experiments~\cite{Gninenko:2014pea,Acciarri:2015uup,Anelli:2015pba} 
can also probe the $X_{3}$ boson through the $\gamma$-$X_{3}$ mixing at loop level, but not for $X_{\pm}$.
The lepton flavor-conserving and -changing scalar bosons can also be searched for through the multi-lepton final states, with the obvious changes in coupling and mass.  
In addition, since some of these scalar bosons are charged under the $SU(2)_{L}$, 
they can be pair produced directly on shell if the collider energy is sufficiently high. 
The signals from the flavor-conserving scalar bosons have shared a common feature 
with the Type-X two-Higgs-doublet model~\cite{Kanemura:2011kx,Kanemura:2012az,Abe:2015oca,Chun:2015hsa}, 
the muon-specific two-Higgs-doublet model~\cite{Abe:2017jqo}, and 
the $A_{4}$ symmetric neutrino mass models~\cite{Ma:2001dn,Fukuyama:2010mz,Fukuyama:2010ff}. 
%

\section{Summary and Discussion\label{sec:summary}}

In this work, we propose a model that extends the Standard Model (SM) gauge group with the $SU(2)_{\mu\tau}$ symmetry that transforms between the second and third generations of the SM leptons.  To break such a symmetry and generate mass splitting between the $\mu$ and $\tau$ leptons, we introduce two exotic scalar fields in addition to the SM Higgs doublet.  One of the new scalar fields is a SM gauge singlet and transforms as a doublet under the new lepton flavor symmetry.  It develops a vacuum expectation value (VEV) much larger than the electroweak scale to completely break the $SU(2)_{\mu\tau}$ symmetry, rendering three degenerate massive gauge bosons denoted by $X_{3,\pm}$.  The other exotic scalar field transforms as an $SU(2)_{\mu\tau}$ triplet and an $SU(2)_L$ doublet simultaneously.  
As a simple scenario, the flavor-conserving component of this $SU(2)_{\mu\tau}$-triplet is induced by  either the $SU(2)_{\mu\tau}$ or SM electroweak symmetry breaking to have a VEV.  With the two sets of Yukawa couplings of $\mu$ and $\tau$ with the SM Higgs doublet field and the exotic $SU(2)_{\mu\tau}$-triplet, one obtains mass splitting between the otherwise degenerate $\mu$ and $\tau$ leptons.

We have computed the contributions of the new particles to the muon anomalous magnetic moment.  For definiteness, we divide the discussions into three scenarios: (i) contributions from the new gauge bosons only, (ii) contributions from the flavor-changing scalar bosons only, and (iii) contributions from the flavor-conserving scalar bosons only.  For each of the scenarios, we have obtained the allowed parameter space for some benchmark parameter sets.  We find that the observed muon $g - 2$ anomaly generally favors mass of the new particles in the regime of ${\cal O}(10-1000)$~GeV and a sufficiently large ratio of the SM Higgs doublet VEV to the exotic triplet VEV.

Since the new gauge bosons and the flavor-violating scalar bosons mediate the $\tau \to \mu \nu_\tau \overline{\nu_\mu}$ decay but not the $\tau \to e \nu_\tau \overline{\nu_e}$ decay, we also discuss the constraint from the measured lepton universality ratio.  We find that if only the new gauge bosons are considered, the lepton universality constraint completely rules out the parameter region favored by the muon $g-2$.  If only the flavor-changing scalar bosons are included, on the other hand, a significantly large parameter space is still allowed by both data.  Current search limits of the charged Higgs boson rule out part of the region.


\bigskip
\bigskip
\section*{Acknowledgments}
CWC would like to thank the hospitality of the Particle Physics Group of Kyoto University, where this work was initiated during his visit.
This research of CWC was supported in part by the Ministry of Science and Technology of Taiwan under Grant No.\ MOST 104-2628-M-002-014-MY4.  %
The work of KT is supported by JSPS Grant-in-Aid for Young Scientists (B) (Grant No. 16K17697), 
by the MEXT Grant-in-Aid for Scientific Research on Innovation Areas (Grant No. 16H00868), 
and by Kyoto University: Supporting Program for Interaction-based Initiative Team Studies (SPIRITS).  
The work of CWC and KT are also supported by JSPS Joint Research Projects (Collaboration, Open Partnership), 
``New Frontier of neutrino mass generation mechanisms via Higgs physics at LHC and flavor physics.''

\appendix
\section{Mass matrices for scalar fields \label{sec:mass-matrices}}
%
\noindent
The mass matrices for the scalar fields associated with the $SU(2)_{\mu\tau}$ symmetry are decomposed as 
\begin{align}
V^{\text{Mass}} = 
%
& \frac12 
\begin{pmatrix} S_{0}^{h} & h_{3}^{} & h_{0}^{} \end{pmatrix}
M_{0,\text{even}}^{2}
\begin{pmatrix} S_{0}^{h} \\ h_{3}^{} \\ h_{0}^{} \end{pmatrix}
%
+\frac12 
\begin{pmatrix} S_{0}^{z} & z_{3}^{} & z_{0}^{} \end{pmatrix}
M_{0,\text{odd}}^{2}
\begin{pmatrix} S_{0}^{z} \\ z_{3}^{} \\ z_{0}^{} \end{pmatrix}
%
+\begin{pmatrix} \omega_{3}^{-} & \omega_{0}^{-} \end{pmatrix}
M_{0\,C}^{2}
\begin{pmatrix} \omega_{3}^{+} \\ \omega_{0}^{+} \end{pmatrix}
\nonumber \\
&
%
+\begin{pmatrix} -S_{-} & \phi_{+}^{0\star} & \phi_{-}^{0} \end{pmatrix}
M_{\pm}^{2}
\begin{pmatrix} -S_{+} \\ \phi_{+}^{0} \\ \phi_{-}^{0\star} \end{pmatrix}
%
+\begin{pmatrix} \phi_{-}^{-} & \phi_{+}^{-} \end{pmatrix} 
M_{\pm\,C}^{2}
\begin{pmatrix} \phi_{+}^{+} \\ \phi_{-}^{+} \end{pmatrix},  
\label{eq:Vmass}
\end{align}
where the electric charges, CP-parity, and lepton flavor property of each set of scalar fields are summarized in Table~\ref{tab:scalarproperty}.

\bigskip
\begin{table}[thb]
\begin{tabular}{ccccc}
\hline\hline
~Original field~ & ~Mass eigenstates~ & ~Electric charge~ & ~~CP~~ & ~Lepton Flavor~ \\
\hline
$(S_0^h, h_3, h_0)$ & $(s, H, h)$ & 0 & even & conserving \\
$(S_0^z, z_3, z_0)$ & $((X_3)_{L}, Z_L, A)$ & 0 & odd & conserving \\
$(\omega_3^+, \omega_0^+)$ & $((W^+)_L, H^+)$ & $+ 1$ & --- & conserving \\
$(S_+, \phi_+^0, \phi_{-}^{0\star})$ & $((X_+)_L, H_+, h_+)$ & 0 & --- & changing \\
$(\phi_{+}^{+}, \phi_{-}^{+})$ & $(\phi_{+}^{+}, \phi_{-}^{+})$ & $+ 1$ & --- & changing \\
\hline\hline
\end{tabular}
\caption{Properties of different sets of scalar fields in the model.  The subscript $L$ in the mass eigenstates denotes the longitudinal mode of the corresponding field.}
\label{tab:scalarproperty}
\end{table}

The mass-squared matrices in Eq.~\eqref{eq:Vmass} are given by
\begin{align}
M_{0,\text{even}}^{2}
=& 
\begin{pmatrix} 
2 \lambda_{S}^{} v_{S}^{2}
& -(2\frac{M^{2}}{v_{S}^{2}} \frac{v_{0}^{2}}{v^{2}} -\lambda_{\Phi S1}^{}) v_{S}^{} v_{3} 
& -(2\frac{M^{2}}{v_{S}^{2}} \frac{v_{3}^{2}}{v^{2}} -\lambda_{0S}^{}) v_{S}^{} v_{0} \\
-(2\frac{M^{2}}{v_{S}^{2}} \frac{v_{0}^{2}}{v^{2}} -\lambda_{\Phi S1}^{}) v_{S}^{} v_{3}
& \frac{M^{2}}{v^{2}} v_{0}^{2} +\lambda_{1} v_{3}^{2}
& -(\frac{M^{2}}{v^{2}} -\lambda_{345}^{}) v_{3} v_{0} \\
-(2\frac{M^{2}}{v_{S}^{2}} \frac{v_{3}^{2}}{v^{2}} -\lambda_{0S}^{}) v_{S}^{} v_{0} 
& -(\frac{M^{2}}{v^{2}} -\lambda_{345}^{}) v_{3} v_{0}
& \frac{M^{2}}{v^{2}} v_{3}^{2} +\lambda_{2} v_{0}^{2}
\end{pmatrix}, \\
M_{0,\text{odd}}^{2}
=& 
( M^{2} -\lambda_{5}^{}v^{2} )
\begin{pmatrix} 0 & 0 & 0 \\ 
0 & v_{0}^{2}/v^{2} & -v_{3}v_{0}/v^{2} \\ 
0 & -v_{3}v_{0}/v^{2} & v_{3}^{2}/v^{2} \end{pmatrix}, 
\\
%
M_{0\,C}^{2}
=& 
\left( M^{2}-\frac{\lambda_{4}+\lambda_{5}}2 v^{2} \right)
\begin{pmatrix} v_{0}^{2}/v^{2} & -v_{3}v_{0}/v^{2} \\ 
-v_{3}v_{0}/v^{2} & v_{3}^{2}/v^{2} \end{pmatrix}, 
\\
M_{\pm}^{2}
=& 
V^{\dag}
\begin{pmatrix} 
0 & 0 & 0 \\
0 & \frac{M^{2}}{v^{2}} v_{0}^{2} \left( 1 + \frac{4v_{3}^{2}}{v_{S}^{2}}\right) & 
-\frac{\lambda_{\Phi S2}}2 v_{S}^{2} \sqrt{1 + 4v_{3}^{2}/v_{S}^{2}} \\
0 & -\frac{\lambda_{\Phi S2}}2 v_{S}^{2} v_{S}^{2} \sqrt{1 + 4v_{3}^{2}/v_{S}^{2}}
& \frac{M^{2}}{v^{2}} v_{0}^{2} -\left( \lambda'''_{1} v_{3}^{2} +\lambda_{5} v_{0}^{2}\right)
\end{pmatrix}
V, \\
M_{\pm\,C}^{2}
=& 
\begin{pmatrix} 
m_{\phi_{}^{\pm}}^{2} 
+ \frac{\lambda_{\Phi S2}}2 v_{S}^{2} - \text{Re}(\kappa)v_{0}v_{3} & 0 \\ 
0 & m_{\phi_{}^{\pm}}^{2} 
- \frac{\lambda_{\Phi S2}}2 v_{S}^{2} + \text{Re}(\kappa)v_{0}v_{3}
\end{pmatrix}, 
\end{align}
where $M^{2}$ is defined as $(M^{2}/v^{2}) v_{3} v_{0} \equiv \lambda_{\mu} v_{S}^{2}/4$,  
and $\lambda_{345} \equiv \lambda_{3}^{} +\lambda_{4} +\lambda_{5}$.
The parameters $\mu_{S}^{2}, \mu_{0}^{2}$ and $m_{\Phi}^{2}$ are eliminated by vacuum conditions. 
The unitary matrix $V$ is given by
\begin{align}
V \equiv 
\begin{pmatrix} 
\frac{v_{S}^{}}{\sqrt{v_{S}^{2}+4v_{3}^{2}}} & \frac{2v_{3}^{}}{\sqrt{v_{S}^{2}+4v_{3}^{2}}}  & 0 \\ 
-\frac{2v_{3}^{}}{\sqrt{v_{S}^{2}+4v_{3}^{2}}} & \frac{v_{S}^{}}{\sqrt{v_{S}^{2}+4v_{3}^{2}}}  & 0 \\ 0 & 0 & 1
\end{pmatrix}
\begin{pmatrix} 
1 & 0 & 0 \\ 0 & \frac1{\sqrt2} & \frac1{\sqrt2} \\ 0 & -\frac1{\sqrt2} & \frac1{\sqrt2} 
\end{pmatrix}. 
\end{align}
%

\section{New gauge interactions for $\mu$ and $\tau$ \label{sec:gauge-int}}
%
\noindent
The gauge kinetic terms for the $\mu$ and $\tau$ leptons are given by  
\begin{align}
i\,\overline{L}_{\alpha} (\cancel{\mathcal{D}})^{\alpha}_{~\beta}{} L^{\beta} 
&= i\,\overline{L_{\mu}} \cancel{D} L_{\mu} + i\,\overline{L_{\tau}} \cancel{D} L_{\tau} 
-\frac{g_{X}^{}}2 X_{3}^{\lambda} (\overline{\mu_{L}^{}} \gamma_{\lambda} \mu_{L}^{}
- \overline{\tau_{L}^{}} \gamma_{\lambda} \tau_{L}^{}
+\overline{\nu_{\mu L}^{}} \gamma_{\lambda} \nu_{\mu L}^{}
- \overline{\nu_{\tau L}^{}} \gamma_{\lambda} \nu_{\tau L}^{}) \nonumber \\
& \qquad 
-\frac{g_{X}^{}}{\sqrt2} [\, 
X_{+}^{\lambda} (\overline{\mu_{L}^{}} \gamma_{\lambda} \tau_{L}^{} 
+\overline{\nu_{\mu L}^{}} \gamma_{\lambda} \nu_{\tau L}^{}) 
+ X_{-}^{\lambda} (\overline{\tau_{L}^{}} \gamma_{\lambda} \mu_{L}^{} 
+\overline{\nu_{\tau L}^{}} \gamma_{\lambda} \nu_{\mu L}^{})], \\
i\,\overline{e_{R}^{}}_{\alpha} (\cancel{\mathcal{D}})^{\alpha}_{~\beta}{} {e_{R}^{}}^{\beta} 
&= i\,\overline{\mu_{R}^{}}_{\mu} \cancel{D} \mu_{R}^{} + i\,\overline{\tau_{R}^{}} \cancel{D} \tau_{R}^{} 
 \nonumber \\
& \qquad 
-\frac{g_{X}^{}}2 X_{3}^{\lambda} (\overline{\mu_{R}^{}} \gamma_{\lambda} \mu_{R}^{}
- \overline{\tau_{R}^{}} \gamma_{\lambda} \tau_{R}^{})
-\frac{g_{X}^{}}{\sqrt2} (X_{+}^{\lambda} \overline{\mu_{R}^{}} \gamma_{\lambda} \tau_{R}^{}
+ X_{-}^{\lambda} \overline{\tau_{R}^{}} \gamma_{\lambda} \mu_{R}^{}), 
\end{align}
where $D^{\lambda}$ is the covariant derivative in the SM.

\section{New Yukawa interactions for $\mu$ and $\tau$ \label{sec:yukawa-int}}
%
\noindent
In this work, we impose the relations $2M^{2}/v_{S}^{2}=\lambda_{\Phi S1}v^{2}/v_{0}^{2}=\lambda_{0S}v^{2}/v_{3}^{2}$ ({\it i.e.}, $\alpha_{1}=\alpha_{2}=0$) for a simplified version of the model. 
In this special case, $\alpha_{3} \equiv \alpha$ is the mixing angle between the CP-even, lepton flavor-conserving Higgs bosons as in the conventional CP-conserving two-Higgs-doublet model.
The flavor-diagonal Yukawa terms rewritten in terms of the mass eigenstates are
\begin{align}
&\overline{L_{\mu}} (y_{0}\,\Phi_{0}+y\,\Phi_{3}) \mu_{R}^{}
+\overline{L_{\tau}} (y_{0}\,\Phi_{0}-y\,\Phi_{3}) \tau_{R}^{} 
+\text{H.c.} \nonumber \\
&= 
+\overline{\mu_{L}^{}} M_{\mu} \mu_{R}^{}
+\overline{\tau_{L}^{}} M_{\tau} \tau_{R}^{} 
\nonumber \\
& \quad 
+ \overline{\mu_{L}^{}} \Big\{ 
\frac{M_{\mu}}{v} \Big( s_{\beta-\alpha} +c_{\beta-\alpha} \frac1{t_{2\beta}^{}}  \Big)
+\frac{M_{\tau}}{v} c_{\beta-\alpha} \frac1{s_{2\beta}^{}} \Big\} \mu_{R}^{} h
-i\, \overline{\mu_{L}^{}} \frac{M_{\mu}}{v} \mu_{R}^{} z
+i\, \sqrt2\, \overline{\nu_{\mu}^{}} \frac{M_{\mu}}{v} \mu_{R}^{} \omega^{+} 
\nonumber \\
& \qquad 
- \overline{\mu_{L}^{}} \Big\{ 
\frac{M_{\mu}}{v} \Big( -c_{\beta-\alpha} + s_{\beta-\alpha}\frac1{t_{2\beta}^{}} \Big)
+\frac{M_{\tau}}{v} s_{\beta-\alpha} \frac1{s_{2\beta}^{}} \Big\} \mu_{R}^{} H
-i\, \overline{\mu_{L}^{}} \Big\{ 
\frac{M_{\mu}}{v} \frac1{t_{2\beta}^{}} 
+\frac{M_{\tau}}{v} \frac1{s_{2\beta}^{}} \Big\} \mu_{R}^{} A
\nonumber \\
& \quad 
+\overline{\tau_{L}^{}} \Big\{ 
\frac{M_{\tau}}{v} \Big( s_{\beta-\alpha} +c_{\beta-\alpha} \frac1{t_{2\beta}^{}}  \Big)
+\frac{M_{\mu}}{v} c_{\beta-\alpha} \frac1{s_{2\beta}^{}} \Big\} \tau_{R}^{} h
-i\,\overline{\tau_{L}^{}} \frac{M_{\tau}}{v} \tau_{R}^{} z
+i\, \sqrt2\, \overline{\nu_{\tau}^{}} \frac{M_{\tau}}{v} \tau_{R}^{} \omega^{+}
\nonumber \\
& \qquad 
-\overline{\tau_{L}^{}} \Big\{ 
\frac{M_{\tau}}{v} \Big( -c_{\beta-\alpha} + s_{\beta-\alpha}\frac1{t_{2\beta}^{}} \Big)
+\frac{M_{\mu}}{v} s_{\beta-\alpha} \frac1{s_{2\beta}^{}} \Big\} \tau_{R}^{} H
-i\, \overline{\tau_{L}^{}} \Big\{ 
\frac{M_{\tau}}{v} \frac1{t_{2\beta}^{}} 
+\frac{M_{\mu}}{v} \frac1{s_{2\beta}^{}} \Big\} \tau_{R}^{} A
\nonumber \\
& \qquad
+i\, \sqrt2\, \overline{\nu_{\mu L}^{}} \Big\{ 
\frac{M_{\mu}}{v} \frac1{t_{2\beta}^{}} 
+\frac{M_{\tau}}{v} \frac1{s_{2\beta}^{}} \Big\} \mu_{R}^{} H^{+}
+i\, \sqrt2\, \overline{\nu_{\tau L}^{}} \Big\{ 
\frac{M_{\tau}}{v} \frac1{t_{2\beta}^{}} 
+\frac{M_{\mu}}{v} \frac1{s_{2\beta}^{}} \Big\} \tau_{R}^{} H^{+}
\nonumber \\
& \qquad
+\text{H.c.},
\end{align}
where 
$1/t_{2\beta}=(1-t_{\beta}^{2})/2t_{\beta}$, $1/s_{2\beta}=(1+t_{\beta}^{2})/2t_{\beta}$, 
and the Yukawa coupling constants are determined by the charged lepton masses as 
\begin{align}
y_{0} = \frac{\sqrt2}{v_{0}} \frac{M_{\mu}+M_{\tau}}2, \quad
y = \frac{\sqrt2}{v_{3}} \frac{M_{\mu}-M_{\tau}}2. 
\end{align}
We note that the requirement of perturbativity for the Yukawa coupling $|y| < 4\pi$ leads to a lower bound on $v_3$:
\begin{align}
v_3 > \frac{|M_{\mu}-M_{\tau}|}{4\sqrt2 \pi} \simeq 90~{\rm MeV} 
\quad 
\Big( \tan\beta \lesssim 2400 \Big).
\end{align}
The flavor off-diagonal Yukawa interactions, on the other hand, are given by 
\begin{align}
&
y\,\sqrt2\,(\overline{L_{\mu}} \Phi_{+} \tau_{R}^{} +\overline{L_{\tau}} \Phi_{-} \mu_{R}^{})
+\text{H.c.} \nonumber \\
&=
\frac{M_{\mu}-M_{\tau}}{\sqrt2\, v c_{\beta}} \Big\{ 
s_{\beta'}\, \overline{\mu}\, \tau\, S_{+}
+\overline{\mu} (c_{\alpha'}c_{\beta'}-s_{\alpha'} \gamma_{5}) \tau H_{+}
-\overline{\mu} (s_{\alpha'}c_{\beta'}+c_{\alpha'} \gamma_{5}) \tau h_{+}
\nonumber \\
& \qquad \qquad \qquad \qquad 
+\sqrt2\, \overline{\nu_{\mu L}^{}} \tau_{R}^{}\, \phi_{+}^{+} 
+\sqrt2\, \overline{\mu_{R}^{}} \nu_{\tau L}^{}\, \phi_{+}^{-} \Big\} 
+\text{H.c.} 
\end{align}

\end{document}